\documentclass{aa}  

\usepackage[varg]{txfonts}
\usepackage{hyperref}
\usepackage{units}
\usepackage{amssymb}
\usepackage[english]{babel}
\usepackage{color}
\usepackage{array}
\usepackage{booktabs}
\usepackage{graphicx}
\usepackage{array}
\usepackage{subfig}
\usepackage{fnpos}
\usepackage{listings}
\usepackage{booktabs}

\usepackage[T1]{fontenc}
\hypersetup{
	colorlinks=true,
	linkcolor=blue,
	filecolor=magenta,      
	urlcolor=blue,
}

\newcolumntype{M}[1]{>{\centering\arraybackslash}m{#1}}
\newcolumntype{N}{@{}m{0pt}@{}}

\newcommand{\HI}{H\textsc{i}}
\newcommand{\Msolar}{M$_{\odot}$}
\newcommand{\kms}{km\,s$^{-1}$}

\begin{document}

\title{Ultra Diffuse Galaxies in clusters\,: the peculiar gas loss of VCC 1964}

\author{R. Taylor\inst{1}
\and V. Partík\inst{1,2}
\and R. Minchin\inst{3}}
\institute{Astronomical Institute of the Czech Academy of Sciences, Prague
\and Astronomical Institute of Charles University, Prague, Czech Republic
\and National Radio Astronomy Observatory\,: Socorro, New Mexico, USA}

\abstract
{Ultra Diffuse Galaxies are low surface brightness systems which have been detected in \HI{} in the field, where their line widths sometimes indicate significant dark matter deficits. They are rarely detected in \HI{} in clusters, making their dynamical properties difficult to assess. The relation between field and cluster populations is unclear.}
{Detecting UDGs entering a cluster could give important clues to their evolution, both in terms of their dynamics but also as to whether they are structurally similar – i.e. if cluster UDGs are generally the same as field UDGs except with less gas and an older stellar population.}
{We use data from two deep Arecibo surveys, the Arecibo Galaxy Environment Survey and the Widefield Arecibo Virgo Environment Survey, to measure the gas content of the UDG-candidate VCC 1964. Optical properties are quantified from the Sloan Digital Sky Survey and DESI Legacy Surveys.}
{We find a significant 9\,kpc offset between the \HI{} and optical components of VCC 1964, no evidence of asymmetry in the \HI{}, and only a modest deficiency level. This suggests a wholesale displacement of the gas content. The line width is 4-5\,$\sigma$ deviant from the baryonic and over 6\,$\sigma$ deviant from the optical forms of the Tully-Fisher relation. The optical component is blue and smooth.}
{VCC 1964 is consistent with a UDG experiencing gas displacement due to ram pressure as it enters the cluster for the first time. Intriguingly, its dynamics imply a significant dark matter deficit, however we cannot rule out that this may be due to the gas being displaced out of equilibrium.}

 
\maketitle
\nolinenumbers

\section{Introduction}
Ultra Diffuse Galaxies (UDGs) are typically defined as being of low surface brightness (fainter than 24 mag\,arcsec$^{-2}$) and large radius (R$_{e}$\,$\ge$\,1.5\,kpc). While their existence has been known for several decades (see \citealt{oldUDGS} for a brief review), only with the discoveries of \cite{vdUDGs} and \cite{kodaUDGs} has it become apparent that these are a distinct sub-class of Low Surface Brightness (LSB) galaxies, present in large numbers in clusters but also now known in groups (\citealt{UDGgroups}) and isolation (\citealt{UDGisolate}). Having diverse properties, with cluster UDGs typically being red and gas-poor compared to those in lower-density environments, multiple pathways for their formation have been suggested e.g. \cite{nudges}. The dynamics of UDGs are particularly intriguing, with observations suggesting that some at least have little or no dark matter (\citealt{DF2, DF4}). While this is still contested, resolved \HI{} gas kinematics have supported this for UDGs found in isolation (e.g. \citealt{UDGisolate}; \citealt{HIUDGSnodm, pina2020}). Although it has been argued that the small rotation velocities suggesting a lack of dark matter are due to errors in estimating the inclination angle (\citealt{wronginc}) or other issues of data quality (\citealt{jones23}), stellar velocity dispersion measurements have confirmed low rotation speeds in at least some objects (\citealt{slowstars}), and statistical studies show that the inclination measurements do not suffer from observational bias (\citealt{nodmnotbias}).

Understanding the relation between cluster and field UDGs is thus important in establishing their more general significance. Importantly, the lower dark matter content has been suggested to result from galaxy-galaxy interactions (e.g. \citealt{silkUDGs}, \citealt{UDGcollisions}), but this scenario has obvious problems for isolated UDGs. If these objects inherently lack dark matter and evolve into cluster-UDGs only through gas loss and quenching of star formation (that is, without any environmental processes depleting their dark matter content), then this points to the existence of a large number of galaxies of reduced dark matter content which pose challenges for cosmological models. Conversely, if field and cluster UDGs are formed by different mechanisms, then the field UDGs without dark matter may represent a rarer, more exotic class of objects. Finding a UDG entering a cluster for the first time, with its gas content not significantly depleted by ram pressure, could give important clues as to the nature of the objects more generally.

Relatively small numbers of field UDGs are known in comparison to cluster members, but far more of the former have dynamical measurements from their gas content. This makes the connection between the two populations unclear. Besides the dynamics, differences in globular cluster (GC) populations have also been noted. \cite{jones23} find that gas-rich UDGs in the field tend to have very few GCs whereas cluster members have rich GC populations. They therefore argue against the results of \cite{UDGX-1correct}, who find that ram pressure stripping can well reproduce the colours and brightness profiles of cluster UDGs if they resulted from an infalling field population. The situation is, however, nuanced. \cite{jones23} note that GCs in UDGs in other clusters differ, with Hydra UDGs having smaller populations, and allow that some small cluster UDGs might result from field infall (we note also that their study did not include Virgo, the focus of \citealt{UDGX-1correct}). \cite{forbes} find evidence for a wide range of GCs within Coma UDGs, with most of their sample being GC-rich but others having populations more consistent with dwarf galaxies. Finally in Virgo, with which the present work is concerned, the situation is unclear. \cite{virgoudgs} find that Virgo UDGs have slightly fewer GCs than elsewhere, but the sample size of 26 is much smaller than in other clusters, and they note that direct comparisons are complicated due to the different luminosities and surface brightnesses of the samples.
	
The number of UDGs with \HI{} detections (hereafter HUD, following \citealt{UDGisolate}) remains low, with \cite{UDGisolate} and \cite{allAAUDGs} finding a total of 252 HUDs; a few more are reported in e.g. \cite{moreHUDs}, \cite{onemoreHUD}, \cite{anotherhud} and \cite{afewmorehuds}, plus 37 in \cite{smudgeshi}. The number of HUDs known in rich clusters is considerably lower. \cite{moreHUDs} identified a total of five HUDs in or near Hickson Compact Groups, while of the 252 ALFALFA-identified HUDs, \cite{allAAUDGs} find only ten within two virial radii of any group (with none identified as being in rich clusters). The latter authors conclude that gas-poor UDGs in clusters may result from the infall and stripping of HUDs, with their optically-diffuse nature not resulting from environmental effects. \cite{fieldUPSB} have recently discovered seven post-starburst UDGs, of which three are isolated and four are satellites of massive systems. Presumably these post-starburst objects should be gas-deficient, however the detection of a UDG actually caught in the process of losing its detected neutral gas is something which has yet to be convincingly demonstrated. 

One such candidate was proposed by \cite{UDGX-1}. They found a candidate UDG NGVS 3543 in the Virgo Cluster, just north of an apparent star-forming wake (with associated \HI{} emission) catalogued as AGC 226178. The two objects were well aligned with the cluster centre, suggestive of ram pressure. This interpretation was challenged by \cite{UDGX-2}, finding that NGVS 3543 is more likely at a distance of only 10\,Mpc, as is the nearby dwarf galaxy VCC 2037. They also discovered a long \HI{} tail extending from both the UDG candidate and patchy stellar emission to the brighter (cluster member) galaxy VCC 2034 – thus they suggest that AGC 226178 is the result of star formation in this stripped gas. Subsequently, \cite{UDGX-1correct} find that their own modelling gives results which are consistent with the \cite{UDGX-2} scenario. Very recently, \cite{UDGX-3} have questioned the existence of the long \HI{} structure, raising further doubts about the origin of AGC 226178.

The complex environment and distance uncertainties of NGVS 3543 make it difficult to assess as a candidate for a UDG experiencing ram pressure stripping. Finding where and how UDGs are quenched remains a matter of some difficulty. \cite{grishin} suggest that UDGs in clusters originate from the gas loss and stellar expansion of relatively massive discs entering clusters on tangential orbits. In contrast, \cite{benudgs} suggest that currently isolated UDGs are backsplash galaxies which were quenched in clusters, but that UDGs were diffuse from their birth in high-spin dark matter halos rather than as a result of evolution within clusters. \cite{fieldUPSB} find that at least some UDGs have been quenched in the field, explicitly ruling out a backsplash origin due to their isolation and post-starburst nature, though this does not preclude quenching also resulting from environmental effects. Finally, \cite{hartke} present the opposite possibility\,: that some UDGs might actually be born in clusters from ram-pressure stripped material (see also \citealt{rpd2}; we will discuss this in more detail in section \ref{sec:discuss}).

Clearly UDGs are difficult to assess as a population in general, and almost certainly result from a multitude of different processes in different environments. In this context, we here present a new candidate for a UDG experiencing gas loss in the Virgo Cluster, VCC 1964. This appears to reside in a far simpler local environment within the Cluster than NGVS 3543. VCC 1964 was first claimed as an \HI{} detection in data from the ALFALFA survey (\citealt{ALFALFA}) in \cite{VCC1964_AA}, but the authors regarded this as very tentative and it was not included in the final ALFALFA data releases. In contrast its detection in the deeper AGES survey (Arecibo Galaxy Environment Survey, see next section for details) was unambiguous, assigned the ID AGESVC1\,275 in \cite{agesvc1}, hereafter AGES\,V. Its significance as a possible UDG was not noted until the SMUDGES project (Systematically Measuring Ultra Diffuse Galaxies; SMDG\,1243207+085703 in \citealt{SMUDGES}), who do not identify the \HI{} counterpart. In this work we present the optical and \HI{} data together. We show that VCC 1964 is a compelling candidate for a UDG losing gas through ram pressure as it enters the cluster. For consistency with previous AGES results, we adopt H$_{0}$\,=\,71\,\kms{}\,Mpc$^{-1}$ and a Virgo cluster distance of 17\,Mpc.

The remainder of this article is organised as follows. In section \ref{sec:obs} we describe the observational data used in this analysis, particularly the Arecibo \HI{} data. Section \ref{sec:results} describes the observational evidence for the two major distinguishing features of VCC 1964\,: its \HI{} gas appears to be displaced from its stellar component, and it has a significant offset from the Tully-Fisher relation. We give a detailed description of how these results were obtained and how secure they are given the uncertainties in the data. Finally we present an interpretation of these findings in section \ref{sec:discuss}.

\section{Observations and data analysis}
\label{sec:obs}
Our \HI{} data comes from both AGES and its direct successor survey WAVES, the Widefield Arecibo Virgo Environment Survey. Both surveys used an identical observing setup, of which full descriptions can be found in \cite{ages}, \cite{isolated}, \cite{agesiv}, and \cite{waves}. We here only summarise the major parameters of the surveys and refer readers to the earlier papers for full technical details. 

\subsection{Observing setup}
\label{sec:obsset}
In brief, AGES was a fully-sampled drift scan survey using the seven beam Arecibo L-band Feed Array instrument ALFA. Each beam records data from two orthogonal polarisations with the final reduced data being an average of both. AGES mapped (among others) two regions of the Virgo cluster to an $rms$ of 0.6\,mJy at spatial and spectral resolutions of 3.5\arcmin{} and 10\,\kms{} (after Hanning smoothing – see below and also section \ref{sec:himeasure}), respectively, with a 1\,$\sigma$ column density sensitivity of 1.5\,$\times$\,10$^{17}$\,cm$^{-2}$$(\Delta\,V\,/\,10)^{1/2}$ (\citealt{AGESM33}). The full bandwidth of AGES was equivalent to a velocity range of -2,000 to +20,000\,\kms{}, however we here only use data pertaining to the Virgo cluster. The main results from Virgo are found in AGES\,V and also AGES\,VI (\citealt{agesvc2}), but the latter covers as different area of the cluster and is not used in the present study.

AGES\,V presented the main AGES results in Virgo, a 10$\times$2 degree field (R.A. versus Declination) centred on M49 deemed AGES\,VC1. This is a complex region with galaxies believed to be in the main body of the cluster (17\,Mpc distance) but also in two infalling clouds at 23 and 32\,Mpc. Distances were assigned based on the measurements and characterisations given in \cite{3DVirgo}. As described in AGES\,V section 4.2 (especially figure 6; see also \citealt{mythesis} chapter 5, section 5.3.2), the boundaries determined for these different groups are uncertain, and the original two-dimensional assignments from \cite{3DVirgo} are, in some cases, not-well matched with the velocity data. However VCC 1964 is well inside the limits of the 17\,Mpc main body of the cluster, being almost three degrees from the boundary. Additionally, as we will discuss in section \ref{sec:tfrcav}, a greater distance would make VCC 1964 an even more extreme object than its data otherwise indicates. If VCC 1964 is a cluster member, this is little room for doubt regarding its distance, but as we shall show, there is somewhat more of a concern as to whether it really is within the Virgo cluster at all.

WAVES was intended as a successor to AGES that would eventually map the entirety of Virgo to the AGES depth and resolution. This was envisaged as a staged project, with the WAVES South data we use here being fully observed and a northern field still in progress at the time the telescope collapsed. As shown in \cite{waves}, the WAVES South field fills in the area immediately north of the AGES\,VC1 field, having the same boundaries in R.A. and a small overlap in declination – exactly where VCC 1964 happens to be situated (AGES\,VC1 has a northern declination limit of +09$^{\circ}$07\arcmin{} whereas the southern limit of WAVES South is +08$^{\circ}$52\arcmin{}). Note that AGES and WAVES use completely different observations, thus they provide independent measurements of the same region in this case. 

The hexagonal beam pattern of ALFA means that at the extreme spatial edges of the surveys, sensitivity is reduced as not all of ALFA's seven beams sample the whole range of declination or R.A. While VCC 1964 is in the region of homogenous sensitivity in the AGES\,VC1 data, it is within the region of reduced sensitivity in the WAVES data. Despite this, WAVES provides two important contributions\,: 1) it verifies the \HI{} parameters as measured in AGES; 2) it enables a search of a larger area for \HI{}-detected galaxies that might be alternative progenitors of AGESVC1\,275. Here we also combine the AGES and WAVES \HI{} data sets, with the overlap region being an average value weighted by the $rms$ levels. First results of the WAVES survey were presented in \cite{waves} while Partík et al. (in preparation) will present the full catalogue of the whole WAVES South region.

\subsection{Data reduction and analysis}
\label{sec:analysis}
Data reduction for both surveys was carried out using a combination of \textsc{livedata}, \textsc{gridzilla} (\citealt{barnes}) and our own post-processing techniques (\citealt{agesvii}); the important aspect of the latter here being the use of a second-order polynomial fit to each spectrum in the data. As standard, the data also has Hanning smoothing applied post-gridding, which removes the effect of Gibbs ringing and decreases the $rms$ (see e.g. \citealt{barnes}). The cost of this is a reduction in spectral resolution by a factor two, not normally a concern\,: the unsmoothed data has a channel width of 5\,\kms{}, so degrading this to 10\,\kms{} does not usually have much impact since this is the approximate width of the \HI{} line itself. Spatial pixels are set to 1\arcmin{}.

Source extraction techniques and efficacy are discussed in detail in \cite{SourceExtraction}. In brief, for both the AGES and WAVES data we have used visual inspection using the \textsc{frelled} package originally developed specifically for AGES data (\citealt{T15, FRELLED5}). This follows our standard two-stage procedure. First, users inspect the data in a 3D volumetric display, interactively masking visible sources (identified through training, based on the known appearance of real extragalactic \HI{} detections – see \cite{SourceExtraction}, in particular section 4.4.1). A volumetric display is extremely convenient for showing the 3D structure of a source, making the masking process rapid and straightforward. The limitation of this is that, due to the necessity to restrict the display of the data to the brightest voxels in 3D (otherwise the display is overwhelmed by noise), only the relatively bright sources are clearly visible. The second stage proceeds with this masked data in 2D, displaying slices of the data (e.g. channel maps) as conventional images. Masking here is more laborious as the user must continually scroll through the images to check exactly where the source is visible, but has the advantage of not requiring any cuts in the data – thus ensuring that even the faintest sources are visible. Once catalogued, as with all AGES and WAVES analysis, we measure the \HI{} parameters using the \textit{msbpect} task from the \textsc{miriad} package (\citealt{miriad}).

The efficacy of the source extraction procedure was quantified in \cite{SourceExtraction}. It was found that the completeness rate closely follows the integrated S/N as defined in \citealt{saint} equation 16, which we give here in the slightly modified form used in most ALFALFA papers (e.g. \citealt{AA100})\,:

\begin{equation}
\label{eqt:aasn}
S/N_{\mathrm{int}} =\frac{1000F_{c}}{W_{50}}\frac{w^{1/2}_{smo}}{rms}
\end{equation}
Where for W50 $<$ 400 km/s, $w_{smo} = W50 / (2 \times v_{res})$, where $v_{res}$ is the velocity resolution in km/s, and for W50 $>$ 400 km/s, $w_{smo} = 400\,km\,s^{-1} / (2 \times v_{res})$. $F_{c}$ is the total flux in Jy\,\kms{}; $rms$ is the $rms$ across the spectrum in mJy.

The completeness rate for visual source extraction of unresolved sources was shown to approach 100\% when the S/N$_{int}$ exceeds a critical threshold of 6.5. For VCC 1964 in the AGES data this was measured at 21.9 – well beyond the point where any follow-up was thought necessary, given that the unusual properties of the source we describe here were not identified during the initial discovery. In section \ref{sec:autosearch} we will also present the results of searching the combined AGES and WAVES data set using automated techniques. The data combination was performed with a custom Python script.

\subsection{Optical data}
\label{sec:optical}
In AGES\,V we relied on optical data from the SDSS and measurements given in the VCC and GOLDMine. Though its LSB-nature was not recognised in AGES\,V, it was noted as unusual due to its morphological assignment (obtained from the GOLDMiNe database of \citealt{goldmine}, itself derived from the Virgo Cluster Catalogue of \citealt{theVCC}) as an early-type galaxy, but its blue colour in the SDSS data ($g$--$i$\,=\,0.32) implied that this classification was erroneous (its position on the colour-magnitude diagram can be seen in AGES\,V figure 14 as the red triangle in the lower right of both panels). Here we use the parameters as measured in SMUDGES, which is based on the more sensitive DESI Legacy Survey (LS) data (\citealt{LS}), with sensitivity of 29\,mag\,arcsec$^{-2}$ ($r$ band, \citealt{LSdepth}). We note that while there is some uncertainty regarding the exact values of the optical parameters (particularly derived quantities such as stellar mass), the qualitative nature of the object is clear\,: this is a blue, structurally smooth LSB galaxy. While its effective radius may be slightly below the putative 1.5\,kpc value for UDG status, this does not affect its physical nature (that is, its properties may be quantitatively different but qualitatively – e.g. dark matter content – similar), as discussed in detail in \cite{nudges}. We note also that alternative, more complex definitions of UDGs have been proposed, e.g. \cite{UDGdefs}, while \cite{fieldUPSB} allow a more liberal definition of UDGs down to R$_{e}$\,$\geq$\,1.0\,kpc. For simplicity, we will default to referring to VCC 1964 as a UDG or UDG-candidate throughout the remainder of this work, noting that it is unclear whether `nearly UDGs' are really a distinct class of object (\citealt{nudges}).

\section{Results}
\label{sec:results}
There are two unusual aspects of VCC 1964 which we will discuss in turn. First, there is a clear offset between the \HI{} and optical components which is highly atypical, with the gas closer in projection to the local cluster centre than the stellar emission. Second, the object is a significant outlier on the Tully-Fisher relation (TFR; we will prefix this with baryonic or optical as appropriate, but the results are the same), having a lower line width than expected given its baryonic mass or optical luminosity. Throughout this section we describe how we obtained these results and the scope for errors, allowing for possible effects of systematic uncertainties (which dominate over the formal imprecision of the  measurements).

\subsection{Measured parameters}   
Table \ref{tab:maintable} gives the object's most significant parameters (both observed and derived) as per our best estimates. For these we use the \HI{} parameters as measured in the combined AGES and WAVES cubes; for the optical data a combination of the SDSS and results from SMUDGES. Specifically\,: the coordinates are those of the \HI{} detection; the apparent magnitudes are from SDSS data using modified SMUDGES parameters; the absolute magnitudes are corrected from Galactic and internal extinction; the stellar mass uses the prescription of \cite{AASDSS}; the inclination angle and effective radius comes from SMUDGES; the rotation velocity accounts for inclination angle as well as correcting for the difference between the maximum versus flat part of the rotation curve.

\begin{table}[h!]
\begin{center}
\caption[bigtable]{Selected properties of VCC 1964 and its associated \HI{} detection, see text for details.}
\label{tab:maintable}
\vspace{-12pt}
\begin{tabular}{c c}\\
\toprule
\textsc{Property} & \textsc{Value(error)}\\
\toprule
R.A. [J2000] & 12:43:17.83(0.7s) \\
Dec. [J2000] & +08:54:53.00(11) \\
Velocity$_{hel}$ & 1,429(2)\,\kms{} \\
Distance & 17\,Mpc \\ 
W50, W20 & 26(2), 41(3)\,\kms{} \\
\HI{} flux & 0.307(0.054)\,Jy\,\kms{} \\
M\HI{} & 2.1(0.4)$\times$10$^{7}$\,\Msolar{} \\
m$_{g}$, m$_{i}$ & 17.61(0.13), 17.27(0.11) \\
M$_{g}$, M$_{i}$ & -13.60(0.13), -13.92(0.11) \\
M$_{*}$ & 2.0$\times$10$^{7}$\,\Msolar{} \\
R$_{e}$ & 17.06(0.41)\arcsec{} \\
R$_{e}$ & 1.41(0.03)\,kpc \\
$\mu_g$ & 24.62(0.01) mag\,arcsec$^{-2}$\\
Inclination & 61(0.7)\,$^{\circ}$ \\
V$_{rot}$ & 14.9(1.6)\,\kms{} \\
M$_{bar}$ & 4.8$\times$10$^{7}$\,\Msolar{} \\
\bottomrule
\end{tabular}
\tablefoot{We do not give estimated uncertainties for stellar or baryonic mass as these would be formally much lower than the systematics. For example different stellar mass recipies give results which differ by a factor of a few (see section \ref{sec:optparams}), even when using the same wavebands.}
\end{center}
\end{table} 

We will discuss all of these in detail in the remainder of this section. One simple parameter is the \HI{} mass, obtained from the flux via the standard equation :
\begin{equation}
\label{eq:mhi}
M\HI{} = 2.36\times10^{5}\times d^{2}\times S_{\HI{}}
\end{equation}
Where \textit{d} is the distance in Mpc. We note that the \HI{} mass is lower than all of the HUDs given in the \cite{allAAUDGs} sample, however they impose a minimum distance of 25\,Mpc; similarly only three objects in the \cite{allAAUDGs} sample have a comparable or more positive (i.e. fainter) absolute $g$ magnitude. To our knowledge, given the other HUD samples cited earlier, this gives VCC 1964 the lowest \HI{} mass of any UDG detected to date. The measured W50  places it in the bottom 10\% of the \cite{allAAUDGs} sample. Colours are difficult to compare given the still limited and inhomogeneous data available for HUDS, but as described in \ref{sec:optical}, VCC 1964 is well within the blue cloud.

The baryonic mass is calculated as M$_{bar}$\,=\,M$_{gas}$\,+\,M$_{*}$. We will discuss the stellar mass separately later in this section. The gas mass is calculated by a correction for helium simply as M$_{gas}$\,=\,1.36\,M\HI{}. We neglect molecular gas as its contribution is likely to be small (e.g. \citealt{dwarfsco}, \cite{BTFRSample}, \citealt{virgodwco}, \citealt{udgsnomolecules}) and would only exacerbate the deviation from the TFR.

\subsection{Offset between the optical and \HI{} components}
\label{sec:offset}
\begin{figure}[h]
\begin{center}
\includegraphics[width=90mm]{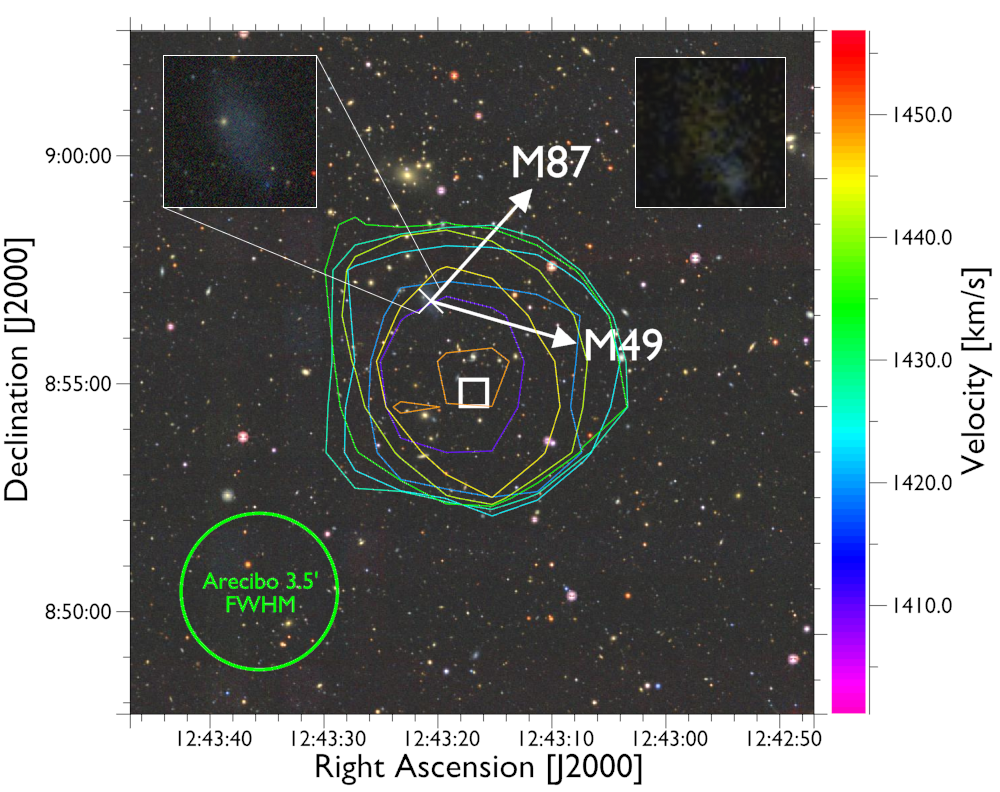}
\caption[ex]{Renzogram of 3.5\,$\sigma$ \HI{} contours of VCC 1964 from the combined AGES/WAVES data sets, overlaid on an LS RGB image. The white square and cross mark the positions of the \HI{} centroid and optical position of the galaxy, respectively. White arrows show the direction vectors to the giant ellipticals M87 and M49, the centre of the main body of the cluster and the local sub-group. The whole image is 15\arcmin{} across, $\approx$\,75\,kpc at the assumed 17\,Mpc distance. The inset images show the same close-up field of view of VCC 1964, optical LS data on the left and GALEX UV on the right.}
\label{fig:main}
\end{center}
\end{figure}

Figure \ref{fig:main} shows the combined AGES and WAVES \HI{} contours overlaid on an LS RGB image. Individually, both \HI{} data sets show the same offset between the \HI{} fitted centroid and the optical coordinates of the galaxy, approximately 1\arcmin{}50\arcsec{}, 9\,kpc in projection. This is a factor 5.5 greater than the AGES median \HI{}-optical offset of 20\arcsec{}, and the similarity of the \HI{} coordinates from two independent data sets strongly suggests that this is a real effect and not due to the effects of noise in fitting the position (with a peak S/N of 19.6 in AGES, noise is not likely to greatly affect the fitted coordinates). The position of the \HI{} and optical counterpart do not well align with the direction to M87 (the nominal centre of the main body of the cluster), with the angular difference between the \HI{}-optical and M87 directional vectors (measured from the optical centre) being 114$^{\circ}$. In contrast the angular difference with the direction of M49 (the centre of the local subgroup) is only 49$^{\circ}$. M87 is 4.63$^{\circ}$ away (1.37\,Mpc in projection) from VCC 1964 while M49 has an angular separation of 3.46$^{\circ}$ (1.03\,Mpc).

As described, we measure the \HI{} parameters using the \textit{msbpect} task from the \textsc{miriad} package. This relies on user-input initial estimates for the spatial and spectral coordinates and the spectral profile range over which to integrate. Although \textit{mbspect} then refines the parameters algorithmically, the subjective nature of the initial guesses cannot be fully removed. We therefore tried adjusting our original input parameters in each case, keeping them visibly consistent with the resultant spectra, and also comparing the fitted spatial position with the \HI{} contours. Adjusting the input parameters and/or using a different data set (AGES\,VC1, WAVES, or the combined cube) can give a fitted position as much as 22\arcsec{} further south than the value reported in table \ref{tab:maintable} (increasing the \HI{} and optical offset), but a position 36\arcsec{} further north is also compatible with the data (lessening the \HI{}/optical offset). The value used here is the compromise case, which gives by far the best visual agreement with the \HI{} contours. Thus we regard the reported offset as secure and not an artifact of data reduction or analysis methodology. 

\subsubsection{Ram pressure as a cause of the gas displacement}
The offset of the \HI{} and optical component is consistent with the \HI{} either being stripped or displaced from (its presumed parent galaxy) VCC 1964 due to ram pressure, a natural explanation given the prevalence of such features in Virgo (\citealt{fading}) – provided, of course, that the object really is a cluster member, which we return to in sections \ref{sec:btfrparamdev} and \ref{sec:tfrcav}. We note also the presence of brighter blue feature at the south-west tip of the optical component, aligned with the gas displacement and also visible in GALEX data. Such a feature is qualitatively consistent with a galaxy experiencing the onset of ram pressure, which has been shown to trigger the formation of molecular gas and an increase in star formation activity (e.g. \citealt{rpsf}, \citealt{almaram}, \citealt{grishin}).

Two caveats potentially complicate this interpretation. Firstly, the misalignment of the optical, \HI{}, and direction vector to M49 of nearly 50$^{\circ}$ implies a rather non-radial orbit. However we believe this is not a significant problem. Observationally, we saw in \cite{fading} that other systems believed to be experiencing ram pressure also show tails collectively pointing away from the local centre but in imperfect alignments (see also \citealt{yagitails, yagibig} and \citealt{grishin}, who interpret misaligned H$\alpha$ tails in the same way). This is likely the result of several contributing factors\,: the cluster is non-spherical and has a complex structure, thus allowing for inherently non-radial orbits; harassment will distort ram pressure tails (in \citealt{kinky} we saw stripped tails of extreme morphological complexity resulting purely from the effects of tidal encounters); density variations within the ICM may cause non-radial tails.
 
Secondly, and more significantly, the \HI{} is found closer to M49 than the optical component. Since the gas should trail the motion of the galaxy, this implies that either this is a backsplash galaxy (having already passed pericentre) or is on a strongly non-radial orbit. 

The backsplash interpretation is in our view problematic. This would require a low-mass dwarf to have traversed at least a couple of megaparsecs through the cluster with only a modest gas loss (see section \ref{sec:gasslossbtfr}) until the present moment, when the entirety of its remaining gas appears to have been displaced. We are not aware of any other backsplash candidates of masses this low which are still relatively gas rich; naively, we would expect a dwarf like this to have been almost completely stripped having crossed such a large swathe of the cluster. While \cite{benudgs} propose backsplash galaxies to explain isolated quiescent UDGs, their objects are red, spheroidal, and gas-poor (perhaps completely gas free, with the bulk of the gas loss occurring at pericentre passage). VCC 1964 is blue, disc-like, and if not obviously gas rich, then it still has gas in its immediate vicinity despite being long past pericentre.

We regard the scenario that the galaxy is on a non-radial orbit but just entering the cluster for the first time as a more promising alternative. This explains the abundance of gas, the alignment of the blue/UV feature with the displaced gas, and the one-sided displacement of the gas from the galaxy. A non-radial orbit would imply (by definition) a significant motion across the sky, which is consistent with the galaxy's small difference in velocity (200\,\kms{}) compared to the cluster itself\,: if indeed the galaxy is losing gas through ram pressure, we would normally expect a much higher motion relative to the ICM than this.

\subsubsection{Gas displacement from a tidal encounter}
We have already suggested tidal encounters as possibly contributing to the angular misalignment of the gas and stellar components. It is also possible that the gas removal itself was largely the result of a tidal encounter rather than ram pressure. We note this as a valid alternative, but a less likely explanation for the gas displacement. VCC 1964 is in a relatively underdense region of the Virgo cluster, with the nearest \HI{} detection 0.82 degrees away (AGESVC1 271, VCC 2033 – 240\,kpc in projection). The nearest VCC galaxy (VCC 1980) is considerably closer at 13.2\arcmin{} (65\,kpc in projection), however this is an early-type galaxy and so unlikely to be the source of the gas. Of course, it could have instead only tidally disturbed VCC 1964 rather than providing its gas content (its systemic velocity is within 200\,\kms{} of that of VCC 1964), but the lack of a counter-tail structure makes this less likely\,: contrary to the effects of ram pressure, we would not expect such a neat, one-sided displacement of the gas in that scenario but rather messier structures (e.g. \citealt{kinky}; a major caveat here is, of course, the unresolved nature of the \HI{} detection). Wholesale gas displacement as apparent here is not unknown in Virgo (e.g. VCC 1249 as discussed in AGES\,V) but is rare.

\subsection{Is the gas really associated with VCC 1964 ?}
\label{sec:cubesearch}
One other possibility for explaining the misaligned gas and stars of VCC 1964 is that the gas is not really associated with the stellar component at all. The proximity of the \HI{} and optical components seem too close to attribute to coincidence, but this has been strongly argued to be the case for the NGVS 3543 system; without an optical redshift we cannot entirely exclude this possibility. It is therefore worthwhile to undertake an additional search of the \HI{} data, as this may reveal other potential sources of the gas.

Given that the \HI{} detection associated with VCC 1964 is a bright source, if VCC 1964 were not itself the donor then we would expect the real source of the gas to be comparably bright. Per the completeness thresholds established in \cite{SourceExtraction}, the chance of missing such a source is negligible. However, VCC 1964 is of a very low line width, and if the parent galaxy had broader emission (as we might expect for more typical galaxies) then it could be much fainter and more likely to have evaded detection. Additionally, there is the possibility of sources which are fainter than VCC 1964 which might give different clues to its origin\,: an alignment of such features might indicate the remains of a stream, or perhaps demonstrate that VCC 1964 is itself in the process of losing gas. Furthermore, the prospect of detecting faint sources in the AGES\,VC1 part of the combined AGES/WAVES data set is relatively high, as this was not previously searched with our more recent source extraction techniques and software described below.

To conduct this search, we extracted a subset cube measuring 112\arcmin{} on a side, about 550\,kpc in projection, spanning the full velocity range of the cluster (0 – 3,000\,\kms{}). This is large enough to include the three nearest \HI{} detections known from our earlier examination. We performed both a visual search and used the automated source finder SoFiA-2 (\citealt{SOFIA2}). The merits of each method are discussed in detail in \cite{SourceExtraction}, but the key point here is that SoFiA is easily able to apply multiple levels of both spatial and spectral smoothing in order to look for emission which is extended on multiple scales. This is typically not done for visual searches owing to the time-consuming nature of the procedure, though in this case we also visually searched the cube after applying several different smoothing kernels. 

\subsubsection{Visual search}
The visual search proceeded in the standard way, as described in section \ref{sec:analysis}. In brief, we visually scanned all R.A.-velocity slices of the cube, interactively masking any emission that resembles a real source (primarily by being sufficiently bright and coherent over several channels/slices). We then examined each candidate source by inspecting the spectra and constructing renzograms. Both of these help by adding an extra check on whether the perceived signal arises from genuine, coherent emission in the data, or if it was only due to a chance alignment of pixels. We explored the use of renzograms in detail in \cite{fading}, demonstrating that real signals typically correspond to coherent emission of two or three beams across (and/or the same number of channels) at a (peak) S/N threshold of 3.5 or greater. Finally, we also inspect optical data of the candidates from the SDSS and LS, as this can provide additional validation for faint sources.

The search was carried out with a conscious effort to be extremely liberal in designating candidate sources, provisionally accepting signals we would usually dismiss as they would be too weak to verify. 14 such candidates were found from a pure visual inspection of the cube. Our additional checks rejected nine of these due to weakness of the signal and incoherency of the spectrum and/or contours (i.e. having contours with different sizes and position from channel to channel), all of which also entirely lacked optical counterparts. The remaining five are marginal but cannot be immediately dismissed for our purposes here. In fact, one of these is a secure detection of the object designated as BC 13 in \cite{rpd2}, which we will describe in detail in a future publication. Stressing that we would ordinarily reject all of the remaining four candidates entirely, none appear to have any obvious relation to VCC 1964. Three are more than 800\,\kms{} different in velocity, BC 13 itself is 28\arcmin{} from VCC 1964, while the remaining object is nearly 1$^{\circ}$ away (300\,kpc in projection).

\subsubsection{Automatic search}
\label{sec:autosearch}
For the SoFiA search, we use the parameters largely as determined in \cite{SourceExtraction}. As above, the chance of detecting even weak signals here is not high, especially given the finding that SoFiA tends to recover the same population of sources that can be found visually. However the overlap between the visual and automatic catalogues is not exact, and SoFiA's capability of spatial smoothing means that there is at least a chance it may detect a larger, extended feature which is not visible in the standard cube.

The search parameters have been shown to closely reproduce, and even slightly exceed, the completeness level of a visual search for unresolved sources. Although this sacrifices reliability, the cube searched here is small, making it feasible to inspect each candidate for validation. We do this in the same way as described above\,: SoFiA candidates are imported as mask regions into our inspection software, which we then use for extracting spectra, generating renzograms, and querying the SDSS and LS (though due to the number of candidates we do not inspect every source by every method). The search parameters are as follows\,:
\begin{lstlisting}
	scfind.kernelsXY = 0
	scfind.kernelsZ = 0, 3, 5, 7, 9, 11, 13, 15, 17, 19, 21, 23, 25, 27, 29, 31, 33, 35, 37, 39, 41
	scfind.threshold = 3.5
	linker.radiusXY = 2
	linker.radiusZ = 3
	linker.minSizeXY = 3
	linker.minSizeZ = 4
	reliability.enable = false
\end{lstlisting}
This returned 113 candidate sources. Of these, apart from those found by other methods and searches in this same area, only two are at all plausible, having weak but clear signals in their spectra, coherent contours in the renzograms, and possible optical counterparts. Furthermore, as with the additional sources found visually, these are still very marginal detections which we would normally disregard from initial publication as they are impossible to confirm without additional observations (we are seeking follow-up for these and other sources in the WAVES area). Moreover, both are $>$\,0.5$^{\circ}$ and $>$\,1,000\,\kms{} from VCC 1964, making them very unlikely to have any association with our main target. 

We also experimented with using different spatial smoothing kernels, which has the potential to reveal faint, extended structures (such as \HI{} bridges and tails) which are below the nominal detection limits. A similar experiment was already performed in \cite{mythesis} (see chapter 7, especially sections 7.3.1 and 7.3.4), which smoothed over 800,000 areas of different sizes and shapes, but only using spatial smoothing. In that earlier study, no credible detections were found, with all candidates being of extremely narrow line width\,: to maintain a line width of a few \kms{} in a stream which is extended for tens of kpc is extremely unlikely, especially given that the line width of the parent galaxies is generally expected to be at least an order of magnitude greater.

For the SoFiA search we tried using spatial smoothing kernels of 9, 11, 13 and 15 pixels, corresponding to physical sizes of 45 to 75\,kpc. All searches resulted in approximately 50 detections (tending to be the same structures in all cases), which have the appearance of randomly-distributed `blobs' of varying velocity widths, distributed throughout the search volume. Based on their position, orientation and morphology, none appear to have any relation to VCC 1964. As a control test we ran the same search on a subset of the Mock spectrometer data used in \cite{SourceExtraction}, where we are certain that no real emission is present due to its high blueshift. The subset was selected to have the same volume as the AGES\,VC1 area we inspect here. 83 sources were found, even more than in the AGES\,VC1 search area, demonstrating that the results are consistent with the effects of noise. Increasing the `scfind.threshold' parameter does not really help, only reducing the number of detections and their sizes.

Based on this, we cautiously suggest that the smoothing used in SoFiA in \cite{rpd1} might be responsible for the claimed detection of a large \HI{} stream in the NGVS 3543 system, which is not seen in either the WAVES or FAST data of similar sensitivity. Given the presence of two clear \HI{} detections in this region, spatial smoothing might increase the apparent size of each source, and since these are relatively close, this may result in the appearance of a connecting bridge. A significant caveat is that the position-velocity diagram used in \cite{rpd1} is not a simple R.A-velocity slice. A more dedicated study is needed to confirm if the large ALFALFA stream is consistent with smoothing of the noise or if the detection does indeed result from a real, physical structure in the data.

In short, few new credible candidate sources were found, and none at all with any apparent relation to VCC 1964. By far the most likely source of the AGESVC1\,275 gas cloud is VCC 1964 itself.

\subsection{Deviation from the Tully-Fisher relation}
\begin{figure}[h]
\begin{center}
\includegraphics[width=90mm]{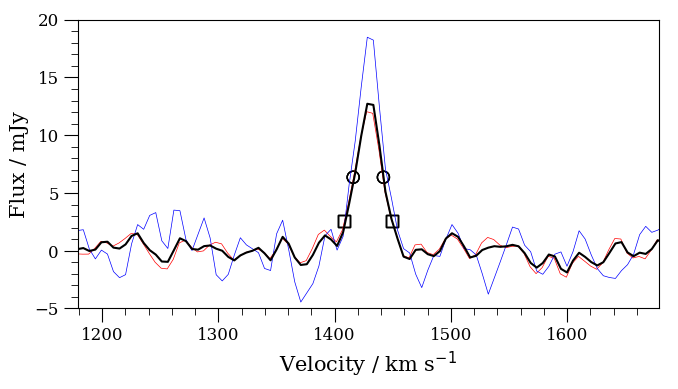}
\caption[ex]{\HI{} spectra of VCC 1964 from AGES (blue), WAVES (red) and the combination (black). Circles and squares show the fitted W50 and W20 velocity widths, respectively.}
\label{fig:spec}
\end{center}
\end{figure}

\begin{figure}[h]
\begin{center}
\includegraphics[width=90mm]{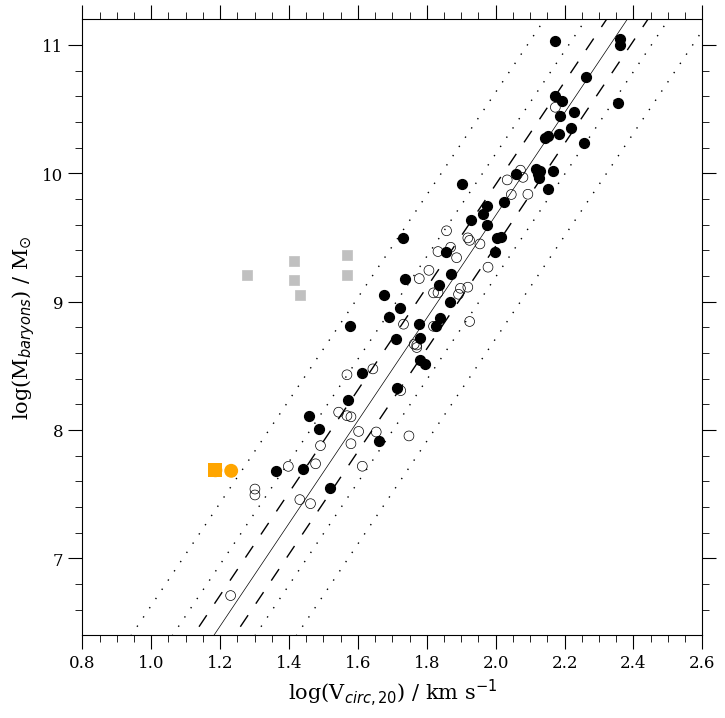}
\caption[ex]{Baryonic Tully-Fisher as in \cite{AGESLeo}. Filled black circles use AGES \HI{} data and SDSS photometry, from the \HI{} detected in the background of the Virgo fields and cluster members with deficiencies $<$\,0.6. Open circles are from table 1 in \cite{BTFRSample}. VCC 1964 is highlighted in orange using the \HI{} data with (circle) and without (square) Hanning smoothing. Filled grey squares show the galaxies in \citealt{HIUDGSnodm}. Outwards from the solid line, the dashed and dotted lines show the 1, 2, and 4\,$\sigma$ scatter.} 
\label{fig:btfr}
\end{center}
\end{figure}

\begin{figure}[h]
\begin{center}
\includegraphics[width=90mm]{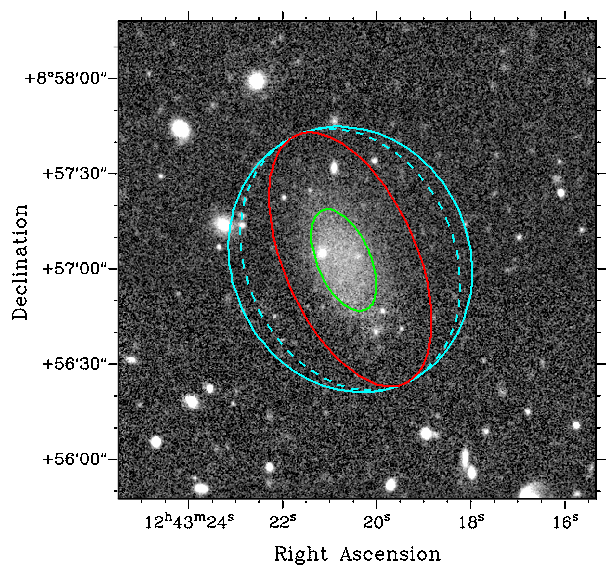}
\caption[ex]{DESI Legacy Survey $g$-band image of VCC 1964 of 2.5\arcmin{} F.O.V., with ellipses showing different inclination estimates. Green is the result from the SMUDGES catalogue. Red shows the aperture we use for photometry on the SDSS data, using the SMUDGES fit but increasing the aperture size by a factor of 2.5 to visually estimate the limit of the galaxy. Cyan ellipses are the same size as the visual fit but with inclination angles which would give a rotation width compatible with the standard (solid) and 2\,$\sigma$-scatter (dashed) of the BTFR.}
\label{fig:opt}
\end{center}
\end{figure}

The line width of the \HI{} detection is among the lowest of the AGES sample with W50\,=\,26\,\kms{}; of the 1,231 published AGES sources only 18 (1.4\%) have line widths equal or lower than this value – spectra are shown here in figure \ref{fig:spec}. This gives the galaxy a 4.5\,$\sigma$ deviation from the Baryonic Tully-Fisher Relation (BTFR; \citealt{BTFR}) as shown in figure \ref{fig:btfr}, increasing to 5.3\,$\sigma$ if we use the unsmoothed AGES cube. This offset is robust to measurement errors in the optical photometry. To bring the galaxy within the 2\,$\sigma$ scatter would require an inclination angle of 40$^{\circ}$, which is grossly incompatible with the optical data (see figure \ref{fig:opt}). Figure \ref{fig:btfr} also shows for comparison the galaxies of \cite{HIUDGSnodm}. These are more than an order of magnitude more massive and generally of higher line width, yet show even stronger deviations. Note that we show these using the values given directly in \cite{HIUDGSnodm} (who use resolved \HI{} maps for modelling the rotation curves), and given the numerous corrections we apply to our own data, the comparison should be taken too literally. We also caution that it is far from the case that every UDG shows similar deviations\,: indeed, some have been claimed to be overmassive rather than having a dark matter deficit (most famously Dragonfly 44, \citealt{DF44_1, DF44_2}, but see also e.g. \cite{vcc1287}, \citealt{forbes}).

Before commenting further on the potential significance of this, we first describe how the BTFR was constructed. We closely follow the procedure given in the appendix of \cite{AGESLeo} (hereafter T22BTFR); for the sake of space, we only summarise the most essential parts of the process here except where we make any alterations. Readers interested in reproducing the BTFR are encouraged to read T22BTFR in full. We note, as in our earlier work, that the \cite{BTFRSample} objects included in figure \ref{fig:btfr} primarily serves as a control sample, demonstrating that the corrections made to our line-width data give results in good agreement with the resolved rotation curves of \cite{BTFRSample} (that sample consists of galaxies with stellar and gas masses ranging approximately from 10$^{6\text{–}9}$\,\Msolar{}, with rotation speeds from 17–150\,\kms{}).

\subsubsection{\HI{} measurements}
\label{sec:himeasure}
We measured the \HI{} parameters of VCC 1964 using four data sets\,: the original AGES\,VC1 data with and without Hanning smoothing, the WAVES cube, and the combined AGES and WAVES data (which we use for the measurements presented here). As noted the WAVES data set is significantly more noisy (1.9\,mJy\,$rms$) than the AGES data in this region and therefore we only use the Hanning smoothed data in this case. Hanning smoothing, as used in standard AGES and WAVES data reduction procedures – indeed this is almost ubiquitous in single-dish \HI{} data – averages half the central channel value with one quarter of the adjacent channels, thus degrading our nominal 5\,\kms{} resolution to 10\,\kms{}. 

For plotting the BTFR, line widths are corrected for spectral resolution (based on the source S/N) and cosmological broadening exactly as in T22BTFR, as are the corrections for rotation (essentially dependent only on inclination) and the difference between the peak and flat parts of the rotation curve (dependent on stellar mass); see section \ref{sec:optparams}. The only difference is that here we have also shown the line width using the unsmoothed data, for which we assume 5\,\kms{} resolution when correcting for instrumental broadening. When using unsmoothed data, we also remeasured the total flux using the same data.

We find there is very little scope for error in the measured flux. Except for the pure WAVES cube, we found the flux did not differ by more than 5\% of the value presented here, with this value being the lowest of those measured. Only the pure WAVES data gives a flux significantly higher (0.411\,Jy\,\kms{}) but this is likely due to the higher noise level, with the effects of noise boosting being well-known (e.g. \citealt{hogg}, \citealt{boris}). Similarly, line widths never vary significantly, being at most 3\,\kms{} less than the reported values in table \ref{tab:maintable} – they were never found to be higher than this.

It is important to note that though the shift on the BTFR using the unsmoothed data is small, it is significant, and we caution that the BTFR deviation is susceptible to relatively small measurement changes. However, the strength of the signal is clearly evident in figure \ref{fig:spec}, and it is unlikely there is any missing broader component that would shift VCC 1964 to higher line widths. We also note that when we remeasured the spectrum as described in section \ref{sec:offset}, the line width measurements never exceeded the value given in table \ref{tab:maintable}. We therefore consider our measured line width an upper limit on the true value. 

\subsubsection{Optical parameters}
\label{sec:optparams}
The optical parameters of the typical galaxies shown in the BTFR presented here are derived using standard SDSS automated photometry. We would prefer to use this for VCC 1964 to ensure homogeneity, but despite being readily visible in the SDSS data, the photometric measurements in this case are clearly inaccurate. For example the SDSS estimate its apparent $g$ magnitude to be 20.16, whereas SMUDGES (which relies on LS data) finds it to be 17.59. Performing aperture photometry on the SDSS data, visually fitting an aperture in \textsc{ds9} without reference to the SMUDGES aperture, we found a very similar value to SMUDGES ($m_{g}$\,=\,17.61). The difference from the SDSS measurement is likely due to the difficulties of algorithmically fitting low surface brightness objects (though we also note that the inclination angle estimates from the SDSS and SMUDGES are virtually identical, with the $r$ band $b/a$ value in the SDSS being 0.50 and the LS value being 0.49, equivalent to 60 and 61$^{\circ}$ respectively).

To avoid using a purely visual fit to the optical data, and ensure we use the same photometric bands for the stellar mass calculation as in T22BTFR (see below), we adopt the following approach : we use the aperture given in the SMUDGES table but use the SDSS $gri$ data for estimating the magnitudes. We expand this aperture by a factor 2.5 to ensure all the light in each band is captured, hence the difference in the sizes of the ellipses in figure \ref{fig:opt}. Point sources visible within the aperture are masked. We use the same ellipses for photometry in the SDSS $g$, $r$ and $i$ bands. With the apparent magnitudes thus obtained, we apply the same extinction and other corrections as described in T22BTFR – these are numerous, and readers interested in the exact result are strongly urged to consult this. 

For comparison, we note that using the aperture purely defined by SMUDGES (including R$_{e}$) would give a $g$ magnitude 0.84 fainter than with our adjusted aperture. However the SMUDGES magnitudes are extrapolated from the Sérsic profile, and our adjusted size for a Sérsic index of $n\,=\,0.51$ (as determined by SMUDGES) is expected to enclose $>$\,98\% of the light. Given the very close agreement between even our visually-estimated aperture and the SMUDGES algorithmic measurements, which used two different data sets, the photometric magnitudes are secure.

A more fundamental problem for the present analysis is the choice of stellar mass prescription. Using the three SDSS bands and the various prescriptions in \cite{bell}, \cite{taylormass}, and \cite{morestellarmasses}, gives a range of possible stellar masses from 5.6$\times$10$^{6}$\,\Msolar{} – 6.4$\times$10$^{7}$\,\Msolar{} \footnote{We provide a simple stellar mass calculator at the following URL : \url{https://photcalc.streamlit.app/}.}. For the value in table \ref{tab:maintable} we choose to give the value as used for the BTFR calculation to ensure consistency. This uses the \cite{taylormass} prescription for the $g$--$i$ colour but includes an additional correction to give stellar masses in agreement with the calculations of \cite{BTFRSample}; again see T22BTFR for full details. While the range of stellar masses calculated is large, reassuringly, the value we have used here is almost exactly the median for our eight calculated values using the various prescriptions. We note that all but the most extreme masses are within a factor two of this value. Due to the systematic issues in calculating the stellar mass, we refrain from commenting on the M\HI{}/M$_{*}$ ratio except to estimate the likely range being from 0.3--4.0.

\subsection{Impact of parameter uncertainty on the BTFR deviation}
\label{sec:btfrparamdev}
As in T22BTFR, we deliberately avoid giving error bars on the BTFR because the many systematic effects are expected to strongly dominate over the formal measurement errors. However, in the main, the deviation of VCC 1964 from the BTFR is not likely subject to any great uncertainties. The velocity width is robust to our choice of data set and profile parameters, with the flux at 20\% of the peak still being well above the noise level, and the integrated flux level giving consistent results (except for the case of WAVES as already discussed, but this would only increase the deviation). We do, however, caution that the small difference in widths measured in the standard and unsmoothed AGES cubes does lead to a noticeable difference in the BTFR, though there is no evidence for a missing broader component. The inclination angle is nearly identical as estimated by the SDSS and SMUDGES, and the angle needed to reconcile the line width with the BTFR is clearly incompatible with the data. 

There are two potential systematic issues which could affect the deviation more strongly (in addition to the distance, discussed earlier). Firstly, as in T22BTFR we have chosen not to correct for turbulence in calculating the rotation width, for the same reason as given therein\,: others have chosen to simply subtract 6.5\,\kms{} from the line width, but this is a rather large correction for low line width galaxies that could introduce considerable additional uncertainties (though it would only increase the deviation from the BTFR).

Secondly, the range of stellar masses allowed by the data is considerable. However in terms of affecting the BTFR the effect of this is likely to be small, as the correction would have to be applied to every galaxy plotted (though, since the extent of this may vary with colour and magnitude, we therefore also show an optical TFR in section \ref{sec:opticalTFR}). More important here is the decision to keep everything self-consistent, so that the deviation can be stated with confidence even if the actual baryonic mass cannot.

We emphasise here that the distance assumption will only have a significant effect if the galaxy is or is not a cluster member. At our assumed distance of 17\,Mpc, the effective radius computed in SMUDGES is (equivalently) just one arcsecond short of the nominal 1.5\,kpc for UDG classification. With the cluster still in the process of assembly, and clouds infalling from 23 and even 32\,Mpc (see AGES\,V for a discussion), a distance underestimate of 1\,Mpc that would increase the physical R$_{e}$ to 1.5\,kpc is certainly plausible; \cite{mei3Dvirgo} estimate the cluster depth at 2.4\,Mpc.

If, on the other hand, VCC 1964 were shown to be well in front of the cluster, then our interpretation would change completely. \cite{UDGX-2} believe the VCC 2037 system is actually at 10\,Mpc distance, low enough to reconcile VCC 1964 with the (2\,$\sigma$) observational scatter in the BTFR. Its effective radius would then be 0.8\,kpc, placing it well short of the usual UDG criterion. However VCC 2037 system is approximately 1.4$^{\circ}$ away in projection, 244\,kpc at that distance, making a direct association with VCC 1964 unlikely. Furthermore, as we have argued, explaining the displacement of the gas and its low line width would become much more difficult in that scenario. In addition, the velocity of VCC 2037 is 300\,\kms{} lower than that of VCC 1964, with VCC 1964 requiring a rather high peculiar velocity of 700\,\kms{} with respect to the cluster if it were actually at 10\,Mpc – and an even lower distance of 5\,Mpc, and so an even higher peculiar velocity, is required to bring VCC 1964 into full agreement with the BTFR.

\subsubsection{Effect of gas loss on the TFR deviation}
\label{sec:gasslossbtfr}
In \cite{agesvc2} we found that strongly deficient galaxies, which have $\leq$\,10\% of the \HI{} content of field galaxies of similar morphology and optical diameter (i.e. deficiencies $>$\,1.0), also show anomalously-low line widths. We attributed this to gas loss truncating the rotation curve such that the detected \HI{} corresponds to the central, steeply-rising part of the curve rather than the outer, flat region. In this scenario the dark matter itself is unaffected\,: instead, the observable \HI{} would simply not probe as great an extent of the rotation curve as for non-deficient galaxies (this is in strong contrast to some ideas in which UDG formation involves actual dark matter depletion, e.g. \citealt{silkUDGs}).

A difficulty with testing this idea here is that the smooth morphology but blue colour of VCC 1964 makes the deficiency calculation uncertain. To explain why, we recall the procedure for calculating deficiency. \HI{} deficiency is defined in \cite{defdef} as :
\begin{equation}
	\label{eqt:hidef}
	\mathrm{DEF}_{\mathrm{HI}} 
	= \log\!\left( M_{\mathrm{HI,\,expected}} \right) 
	- \log\!\left( M_{\mathrm{HI,\,observed}} \right)
\end{equation}
Since this is a logarithmic quantity, a deficiency of 1.0 means a galaxy has lost 90\% of its original gas content, a deficiency of 2.0 implies a loss of 99\%, etc. The expected gas content is defined as follows\,:
\begin{equation}
	M_{\mathrm{HI,\,expected}} = a + b\,log(d) 
\end{equation} 

Where $a$ and $b$ are coefficients derived from a sample of isolated field galaxies (where environmental effects are believed to be minimal). Traditionally these are defined based on the Hubble-type morphology of the galaxy in combination with its optical diameter in kpc ($d$); the problem with VCC 1964 being that its smooth structure but blue colour make it uncertain which values of $a$ and $b$ we should adopt (this was why it was excluded from the TFR plots in our earlier papers). In addition, various studies have come up with different values for these coefficients. However, using the same parameters (of \citealt{solanes} for calculating deficiency as we used in \citealt{agesvc2}), we calculate deficiency to be 0.73 if using the generalised values and 0.48 if a dwarf irregular or later-type (i.e. the galaxy retains 19 or 33\% of its expected original gas content, respectively)\footnote{We provide a simple online deficiency calculator, with $a,\;b$ coefficients from multiple sources, at the following URL : \url{https://rhysyt-hicalculators-hidefcal-qqnsnn.streamlit.app/}.}. 

Thus, comparing to our previous findings, it does not appear likely that VCC 1964 has lost sufficient gas to truncate its rotation curve (suggesting its dynamics are indeed intrinsically unusual) – with the important caveat that the stripping may still have affected its line width by perturbing the object from equilibrium, rather than truncating its rotation curve. We stress here that regardless of the cause, deviations from the BTFR (even in Virgo) are rare among `normal' galaxies (with the dynamics of UDGs still being under dispute). Such similar deviations as we saw in Virgo were only for a very few, much brighter galaxies than VCC 1964 of extreme deficiency, and we saw no evidence of any correlation with deficiency with BTFR offset. Whatever the cause of the deviation it must be an unusual mechanism. On the other hand, VCC 1964 clearly is an unusual object, and there are few if any other objects which allow for comparisons. We will return to this point in the Discussion.

\subsection{Alternative, optical TFR}
\label{sec:opticalTFR}
While the deviation from the computed BTFR appears robust to measurement errors and stellar mass estimates, as an additional test we also construct an optical version of the TFR. We follow the prescription of \cite{opticalTFR} (K20), who calibrate the TFR using over 10,000 spiral galaxies with SDSS, WISE, and ALFALFA data. Using an optical form of the TFR has the advantage of avoiding the large variations possible from stellar mass recipes, and the K20 methodology also includes a more sophisticated correction (as opposed to linearly subtracting a constant value) for turbulence in the line width. Below we give only a brief summary of our corrections to ensure reproducibility; readers interested in the physical basis of this approach should consult K20.

The optical TFR is constructed as follows. The correction for Galactic extinction is done in the same way as for the BTFR. Lacking the WISE photometry used in K20, we follow their colour-adjusted method. First, after correcting for Galactic extinction and applying a K-correction, apparent magnitudes are converted to pseudo-magnitudes as per their equations 16–18\,:
\[\Delta M_{g} = 0.73 \left( m_{g,e,k} - m_{i,e,k} \right) - 0.45\]
\[C_{g} = m_{g,e,k} - \Delta M_{g}\]
Where $C_{g}$ is a pseudo-magnitude that in effect corrects for internal extinction, converted to an absolute magnitude using the standard procedure.

\begin{figure}[h]
\begin{center}
\includegraphics[width=90mm]{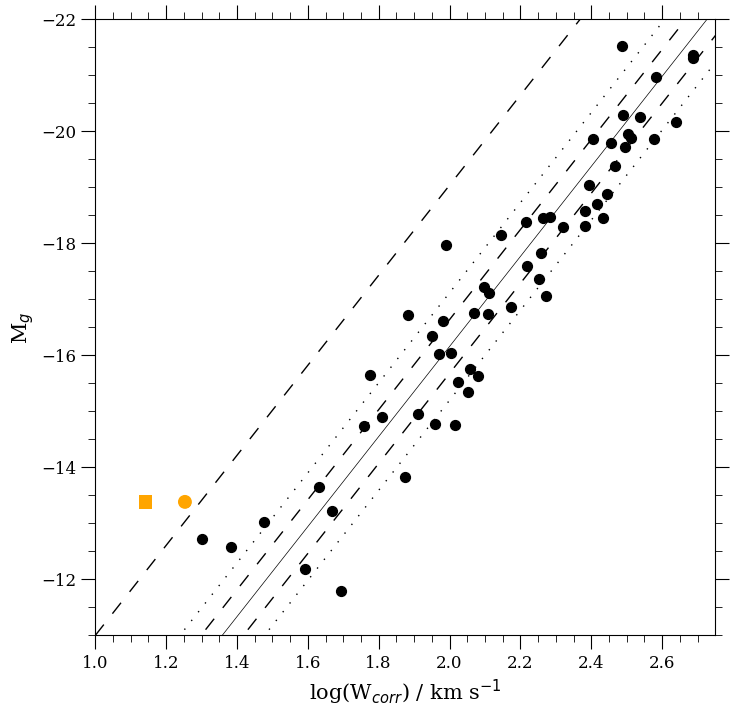}
\caption[ex]{Optical form of the TFR following the methodology of K20. Black points are our AGES galaxies, with VCC 1964 highlighted in orange (as in figure \ref{fig:btfr} the orange square uses \HI{} data without Hanning smoothing). The solid line is the relation obtained in K20 with the dashed and dotted lines showing their estimated 1 and 2\,$\sigma$ scatter. The upper dashed line shows the 6\,$\sigma$ deviation.}
\label{fig:classicTFR}
\end{center}
\end{figure}

Line widths are corrected as follows (following the procedure of \cite{hidatabase}, as adopted in K20; we give the equations here in their formalism). Our W20 values, measured using the peak flux, are converted to the W50 range that encloses 90\% of the total \HI{} flux :
\[W_{m50} = (W20 - 17.7) / 0.988  \]
These are corrected for redshift and resolution as follows :
\[W_{corr} = (W_{m,50} / (1 + z)) - 2.0\,v_{res}\,0.25 \]
The correction for turbulence is given by :
\[x = W_{m,50} / k_{1} \]
\[p = e^{-x^{2}} \]
\[W_{mx}^{2} = W_{m,50}^{2} + k_{2}^{2}(1 - 2\,p) - 2\,W_{m,50}\,k_{2}\,(1 - p)  \]
Where $k_{1}$ and $k_{2}$ are the constants 100\,\kms{} and 9\,\kms{} respectively.
Finally we correct for inclination using the same values as in the BTFR, using the full line width (rather than half) for consistency with K20. We plot their optical TFR :
\[M_{g} = -8.04*(\mathrm{log}(W_{mx} - 2.5) - 20.18 \]
Where M$_{g}$ is the pseudo-absolute-magnitude, with K20 giving a 1\,$\sigma$ scatter in this relation of 0.48. Our result is shown in figure \ref{fig:classicTFR}. The deviation of VCC 1964 increases significantly to over 6\,$\sigma$. We submit that while the physical interpretation of this is certainly an open question (primarily the issue is whether this reflects the intrinsic dynamics of the galaxy or is a result of the gas – quite literally – being pushed out of equilibrium), the deviation itself is clear.

\subsection{Other TFR caveats}
\label{sec:tfrcav}
The final issue relevant for understanding this object is the unresolved nature of the \HI{} detection. A resolved rotation curve would show whether the \HI{} line width is really probing the flat part of the curve or if this has perhaps been truncated (or otherwise affected) by the assumed ram pressure stripping. Additionally, since we can only constrain the \HI{} size to be less than the Arecibo beam size (17\,kpc at the Virgo distance), we cannot give an estimate of the dynamical mass with any meaningful precision. If the rotation curve is flat then the deviation from the BTFR indicates a deficit of dark matter compared to typical galaxies, but without a radius for the \HI{}, we cannot say how much. For comparison, if the line width were within the 2\,$\sigma$ BTFR scatter or in perfect agreement with the \cite{BTFRSample} relation, this would imply a dynamical mass 2--4 times greater than its present value. More generally, a resolved \HI{} map could show if the gas has ordered motions or if the stripping has disturbed it from equilibrium, which could invalidate the BTFR interpretations. Numerical modelling could also be used to investigate if and how the gas rotation changes when displaced from its parent galaxy.

We have discussed the possibility of an inaccurate distance estimate in section \ref{sec:btfrparamdev}, noting that 10\,Mpc – the same as \cite{UDGX-2} measured for NGVS 3543, and also for VCC 2037 in \citealt{VCC2037} – would bring VCC 1964 within the observed scatter in the BTFR. We find this unlikely due to the high peculiar velocity this would imply for VCC 1964, its large separation from the system described in \cite{UDGX-2}, and the lack of an obvious cause of gas removal in this scenario. We add here two additional points. First, \cite{UDGX-2} identified a much larger \HI{} stream as the source of the gas and there were multiple optical structures in close proximity to NGVS 3543. No such features are apparent around VCC 1964. Furthermore the prospect of a second candidate UDG projected against Virgo, with a similarly adjacent but actually unrelated gas cloud, is highly improbable. Of course, only a robust distance measurement to VCC 1964 could fully settle the dilemma, though we note that such estimates are sometimes far from clear-cut – see the introduction of \cite{distanceproblems} for a summary of a particularly well-known example. Second, while it is possible that VCC 1964 is actually at a larger distance (say within the 23 or 32\,Mpc) cloud, this would not change the estimate line width but would increase its inferred baryonic mass, thus only making the BTRF-discrepancy more pronounced. 

In short, while there are legitimate questions as to the precise numerical values for all parameters, there is little indication that this could change any of the major conclusions presented here. Uncertainties of the velocity width are small, and the estimated width is likely an overestimate due to turbulence (resolved \HI{} observations would greatly help with this). The inclination angle is consistent between two independent data sets and incompatible with bringing the line width into agreement with the BTFR. The offset of the optical and \HI{} position is robust in two independent data sets, albeit again better resolution would be of benefit here. The stellar mass estimate is subject to a larger uncertainty, but this would affect other galaxies on the BTFR so is very unlikely to explain the deviation of VCC 1964. 

\section{Discussion}
\label{sec:discuss}
VCC 1964 is a candidate for a UDG losing gas in the Virgo cluster due to ram pressure stripping. The displacement of its gas and stellar component align well with the vector towards M49, the centre of the local maximum of X-ray gas. Conversely, there are no other plausible sources of the \HI{} gas within 250\,kpc and the simple offset from the optical counterpart does not match expectations from a tidal encounter\,: in that case we (typically) would expect the gas to be asymmetrically disturbed around the centre of mass, rather than displaced wholesale. 

We note there is one other candidate for a Virgo Cluster UDG currently experiencing the onset of ram pressure. \cite{UDGX-1correct} select a sample of low surface brightness galaxies from the Next Generation Virgo Survey, of which five are detected in \HI{} from ALFALFA. They state that this does not include any of their 26 LSB galaxies identified as UDGs, but they define UDGs using the more complex criteria of \cite{virgoudgs}. Using the standard effective radius and surface brightness criteria (see \citealt{virgoudgs} table 2), two of their \HI{} detections would be classified as UDGS\,: VCC 169 and AGC 220597. They note the former as possibly suggestive of the onset of ram pressure stripping based on its H$\alpha$ detection. This object is also detected in WAVES. We will comment further on this in the final WAVES catalogue paper, but note here that it does not show any offset between the \HI{} and optical positions, nor any disturbances in its \HI{} contours; it may be at an earlier stage of stripping than VCC 1964. The highly irregular optical morphology of the object makes any estimate of its inclination angle difficult, and thus we cannot comment on its position on the BTFR.

Recently, \cite{rpd2} have built on the work of \cite{rpd1} to describe a class of stellar systems in Virgo they dub `blue blobs'. They interpret these as ram pressure dwarfs, analogous to tidal dwarfs but with a ram pressure origin of the gas rather than tidal interactions. Numerical simulations in \cite{blobsim} show that such objects could be gravitationally self-bound and stable on long ($>$\,1\,Gyr) timescales. If in the cluster, VCC 1964 almost certainly has a stellar mass considerably greater than most of the known blue blobs (median 1.3$\times$10$^{5}$\,\Msolar{} from the catalogue of \citealt{rpd2}), but the same principle could apply, with its stellar mass being recently formed within the \HI{}. On the other hand, the smooth, symmetrical nature of VCC 1964 would not be consistent with this, nor would the offset between the \HI{} and the optical components.

The offset of VCC 1964 from the BTFR can be reconciled by a distance of 10\,Mpc, the same as measured for VCC 2034 (\citealt{UDGX-2}). However the lack of nearby galaxies would then make its displaced gas difficult to explain. If in Virgo, the offset from the BTFR is also not necessarily problematic, as this has been shown to be the case for many other UDGs. Detailed hydrodynamic studies are needed to show if such a galaxy would retain its very narrow \HI{} line width – or indeed if the removal is itself the cause of the narrow line – and also to address whether the stellar disc would remain undisturbed (as it appears to be) during the removal of the relatively massive gas component. This is particularly significant given its apparent deficit of dark matter.

The interpretation of the offset of VCC 1964 is much the most uncertain aspect of our analysis. We have previously shown examples of objects with similar offsets due to a truncation of the \HI{} disc (i.e. the measured line width), but the deficiency of this object is not consistent with this. Despite this, the gas may not be in stable rotation as the BTFR assumes, and indeed that the gas is displaced is certainly evidence of this. The issue here is the great deal of uncertainty regarding what happens to the line width of stripped gas. Many objects show a truncation of the rotation curve of the remaining gas but with significantly larger widths in the stripped material, e.g. \cite{tailwidth3, tailwidth2, tailwidth1}; in simulations of tidal stripping, \cite{kinky} find that stripped gas tends to increase in line width. Of most direct relevance, \cite{tailwidth4} find that the velocity dispersion of stripped H$\alpha$ tails increases due to turbulent mixing with the ICM. This implies that the material remaining in the disc, despite its presently lower line width, would increase in width on its removal into the ICM.

Since the gas appears to have been wholly removed from the disc of VCC 1964, the above could be taken to indicate that we are, if anything, overestimating its line width – but there are many caveats. The tail widths in \cite{tailwidth4} increase  on scales of 20\,kpc whereas the gas of VCC 1964 is only offset by 9\,kpc. More significantly, the \cite{tailwidth4} tails show strong local variations in velocity dispersion, not a smooth linear increase. The situation of VCC 1964 is also different in that we appear to be dealing far more with gas displacement than continuous deformation into an elongated tail, and without similar galaxies for comparison (and/or detailed hydrodynamical modelling) the expected change in line width is far from obvious. 
	
We have also described how UDGs differ substantially in the GC populations, and thus it is hard to generalise how this one particular example relates to UDGs in clusters more widely. Indeed, its very peculiarities are what make this a remarkable system. It has the lowest \HI{} mass detected in a UDG to date, displaced from its stellar component though the optical emission is smooth, with the \HI{} detected closer to the local cluster centre than the stars. Explaining this collective strangeness, together with its TFR offset, remains challenging.

A direct distance measurement is needed to determine if VCC 1964 really is a UDG entering the Virgo cluster for the first time. If so, this would suggest that gas-depleted cluster UDGs – the majority found so far – can form from the gas-rich population known in the field,  implying a common (though as yet poorly constrained) origin for at least some members of both types of object. Of course, VCC 1964 is itself just one object, and it is conceivable that multiple formation pathways may still be at work. Only larger sample sizes could address this issue and help understand the formation of similar objects apparently lacking in dark matter.

\section*{Data availability}
The AGESVC1 data cube is available in its entirety on request to the corresponding author. The full WAVES cube is awaiting publication, but the combined AGES-WAVES data set used here can be similarly made available on request. All other data sources used in this work are public.

\begin{acknowledgements}
We thank our numerous colleagues at the ASU for their helpful and productive discussions. We also thank the anonymous referee whose comments have significantly improved the manuscript. 

This work was supported by the institutional project RVO:67985815, the Czech Ministry of Education, Youth and Sports from the large infrastructures for Research, Experimental Development and Innovations project LM 2015067, and the Charles University project GA UK No. 376425. 

This research has made use of the SDSS. Funding for the Sloan Digital Sky Survey V has been provided by the Alfred P. Sloan Foundation, the Heising-Simons Foundation, the National Science Foundation, and the Participating Institutions. SDSS acknowledges support and resources from the Center for High-Performance Computing at the University of Utah. SDSS telescopes are located at Apache Point Observatory, funded by the Astrophysical Research Consortium and operated by New Mexico State University, and at Las Campanas Observatory, operated by the Carnegie Institution for Science. The SDSS web site is www.sdss.org. SDSS is managed by the Astrophysical Research Consortium for the Participating Institutions of the SDSS Collaboration, including the Carnegie Institution for Science, Chilean National Time Allocation Committee (CNTAC) ratified researchers, Caltech, the Gotham Participation Group, Harvard University, Heidelberg University, The Flatiron Institute,  The Johns Hopkins University, L’Ecole polytechnique fédérale de Lausanne (EPFL), Leibniz-Institut für Astrophysik Potsdam (AIP), Max-Planck-Institut für Astronomie (MPIA Heidelberg), Max-Planck-Institut für Extraterrestrische Physik (MPE), Nanjing University, National Astronomical Observatories of China (NAOC), New Mexico State University, The Ohio State University, Pennsylvania State University, Smithsonian Astrophysical Observatory, Space Telescope Science Institute (STScI), the Stellar Astrophysics Participation Group, Universidad Nacional Autónoma de México (UNAM), University of Arizona, University of Colorado Boulder, University of Illinois at Urbana-Champaign, University of Toronto, University of Utah, University of Virginia, Yale University, and Yunnan University.
\end{acknowledgements}

\bibliographystyle{aa}
\bibliography{references}	

@ARTICLE{oldUDGS,
       author = {{Conselice}, Christopher J.},
        title = "{Ultra-diffuse Galaxies Are a Subset of Cluster Dwarf Elliptical/Spheroidal Galaxies}",
      journal = {RNAAS},
         year = 2018,
        month = mar,
       volume = {2},
       number = {1},
          eid = {43},
        pages = {43},
          doi = {10.3847/2515-5172/aab7f6},
archivePrefix = {arXiv},
       eprint = {1803.06927},
 primaryClass = {astro-ph.GA},
       adsurl = {https://ui.adsabs.harvard.edu/abs/2018RNAAS...2...43C}
}

@ARTICLE{UDGdefs,
       author = {{Lim}, Sungsoon and {C{\^o}t{\'e}}, Patrick and {Peng}, Eric W. and {Ferrarese}, Laura and {Roediger}, Joel C. and {Durrell}, Patrick R. and {Mihos}, J. Christopher and {Wang}, Kaixiang and {Gwyn}, S.~D.~J. and {Cuillandre}, Jean-Charles and {Liu}, Chengze and {S{\'a}nchez-Janssen}, Rub{\'e}n and {Toloba}, Elisa and {Sales}, Laura V. and {Guhathakurta}, Puragra and {Lan{\c{c}}on}, Ariane and {Puzia}, Thomas H.},
        title = "{The Next Generation Virgo Cluster Survey (NGVS). XXX. Ultra-diffuse Galaxies and Their Globular Cluster Systems}",
      journal = {\apj},
         year = 2020,
        month = aug,
       volume = {899},
       number = {1},
          eid = {69},
        pages = {69},
          doi = {10.3847/1538-4357/aba433},
archivePrefix = {arXiv},
       eprint = {2007.10565},
 primaryClass = {astro-ph.GA},
       adsurl = {https://ui.adsabs.harvard.edu/abs/2020ApJ...899...69L}
}

@ARTICLE{vdUDGs,
       author = {{van Dokkum}, Pieter G. and {Abraham}, Roberto and {Merritt}, Allison and {Zhang}, Jielai and {Geha}, Marla and {Conroy}, Charlie},
        title = "{Forty-seven Milky Way-sized, Extremely Diffuse Galaxies in the Coma Cluster}",
      journal = {\apjl},
         year = 2015,
        month = jan,
       volume = {798},
       number = {2},
          eid = {L45},
        pages = {L45},
          doi = {10.1088/2041-8205/798/2/L45},
archivePrefix = {arXiv},
       eprint = {1410.8141},
 primaryClass = {astro-ph.GA},
       adsurl = {https://ui.adsabs.harvard.edu/abs/2015ApJ...798L..45V}
}

@ARTICLE{kodaUDGs,
       author = {{Koda}, Jin and {Yagi}, Masafumi and {Yamanoi}, Hitomi and {Komiyama}, Yutaka},
        title = "{Approximately a Thousand Ultra-diffuse Galaxies in the Coma Cluster}",
      journal = {\apjl},
         year = 2015,
        month = jul,
       volume = {807},
       number = {1},
          eid = {L2},
        pages = {L2},
          doi = {10.1088/2041-8205/807/1/L2},
archivePrefix = {arXiv},
       eprint = {1506.01712},
 primaryClass = {astro-ph.GA},
       adsurl = {https://ui.adsabs.harvard.edu/abs/2015ApJ...807L...2K}
}

@ARTICLE{UDGgroups,
       author = {{Rom{\'a}n}, Javier and {Trujillo}, Ignacio},
        title = "{Ultra-diffuse galaxies outside clusters: clues to their formation and evolution}",
      journal = {\mnras},
         year = 2017,
        month = jul,
       volume = {468},
       number = {4},
        pages = {4039-4047},
          doi = {10.1093/mnras/stx694},
archivePrefix = {arXiv},
       eprint = {1610.08980},
 primaryClass = {astro-ph.GA},
       adsurl = {https://ui.adsabs.harvard.edu/abs/2017MNRAS.468.4039R}
}

@ARTICLE{UDGisolate,
       author = {{Leisman}, Lukas and {Haynes}, Martha P. and {Janowiecki}, Steven and {Hallenbeck}, Gregory and {J{\'o}zsa}, Gyula and {Giovanelli}, Riccardo and {Adams}, Elizabeth A.~K. and {Bernal Neira}, David and {Cannon}, John M. and {Janesh}, William F. and {Rhode}, Katherine L. and {Salzer}, John J.},
        title = "{(Almost) Dark Galaxies in the ALFALFA Survey: Isolated H I-bearing Ultra-diffuse Galaxies}",
      journal = {\apj},
         year = 2017,
        month = jun,
       volume = {842},
       number = {2},
          eid = {133},
        pages = {133},
          doi = {10.3847/1538-4357/aa7575},
archivePrefix = {arXiv},
       eprint = {1703.05293},
 primaryClass = {astro-ph.GA},
       adsurl = {https://ui.adsabs.harvard.edu/abs/2017ApJ...842..133L}
}

@ARTICLE{nudges,
       author = {{Buzzo}, Maria Luisa and {Forbes}, Duncan A. and {Jarrett}, Thomas H. and {Marleau}, Francine R. and {Duc}, Pierre-Alain and {Brodie}, Jean P. and {Romanowsky}, Aaron J. and {Ferr{\'e}-Mateu}, Anna and {Hilker}, Michael and {Gannon}, Jonah S. and {Pfeffer}, Joel and {Haacke}, Lydia},
        title = "{The multiple classes of ultra-diffuse galaxies: can we tell them apart?<SUP></SUP>}",
      journal = {\mnras},
         year = 2025,
        month = jan,
       volume = {536},
       number = {3},
        pages = {2536-2557},
          doi = {10.1093/mnras/stae2700},
archivePrefix = {arXiv},
       eprint = {2412.01901},
 primaryClass = {astro-ph.GA},
       adsurl = {https://ui.adsabs.harvard.edu/abs/2025MNRAS.536.2536B}
}

@ARTICLE{DF2,
       author = {{van Dokkum}, Pieter and {Danieli}, Shany and {Cohen}, Yotam and {Merritt}, Allison and {Romanowsky}, Aaron J. and {Abraham}, Roberto and {Brodie}, Jean and {Conroy}, Charlie and {Lokhorst}, Deborah and {Mowla}, Lamiya and {O'Sullivan}, Ewan and {Zhang}, Jielai},
        title = "{A galaxy lacking dark matter}",
      journal = {\nat},
         year = 2018,
        month = mar,
       volume = {555},
       number = {7698},
        pages = {629-632},
          doi = {10.1038/nature25767},
archivePrefix = {arXiv},
       eprint = {1803.10237},
 primaryClass = {astro-ph.GA},
       adsurl = {https://ui.adsabs.harvard.edu/abs/2018Natur.555..629V}
}

@ARTICLE{DF4,
       author = {{van Dokkum}, Pieter and {Danieli}, Shany and {Abraham}, Roberto and {Conroy}, Charlie and {Romanowsky}, Aaron J.},
        title = "{A Second Galaxy Missing Dark Matter in the NGC 1052 Group}",
      journal = {\apjl},
         year = 2019,
        month = mar,
       volume = {874},
       number = {1},
          eid = {L5},
        pages = {L5},
          doi = {10.3847/2041-8213/ab0d92},
archivePrefix = {arXiv},
       eprint = {1901.05973},
 primaryClass = {astro-ph.GA},
       adsurl = {https://ui.adsabs.harvard.edu/abs/2019ApJ...874L...5V}
}

@ARTICLE{HIUDGSnodm,
       author = {{Mancera Pi{\~n}a}, Pavel E. and {Fraternali}, Filippo and {Adams}, Elizabeth A.~K. and {Marasco}, Antonino and {Oosterloo}, Tom and {Oman}, Kyle A. and {Leisman}, Lukas and {di Teodoro}, Enrico M. and {Posti}, Lorenzo and {Battipaglia}, Michael and {Cannon}, John M. and {Gault}, Lexi and {Haynes}, Martha P. and {Janowiecki}, Steven and {McAllan}, Elizabeth and {Pagel}, Hannah J. and {Reiter}, Kameron and {Rhode}, Katherine L. and {Salzer}, John J. and {Smith}, Nicholas J.},
        title = "{Off the Baryonic Tully-Fisher Relation: A Population of Baryon-dominated Ultra-diffuse Galaxies}",
      journal = {\apjl},
         year = 2019,
        month = oct,
       volume = {883},
       number = {2},
          eid = {L33},
        pages = {L33},
          doi = {10.3847/2041-8213/ab40c7},
archivePrefix = {arXiv},
       eprint = {1909.01363},
 primaryClass = {astro-ph.GA},
       adsurl = {https://ui.adsabs.harvard.edu/abs/2019ApJ...883L..33M}
}

@ARTICLE{nodmnotbias,
       author = {{Hu}, Hui-Jie and {Guo}, Qi and {Zheng}, Zheng and {Yang}, Hang and {Tsai}, Chao-Wei and {Zhang}, Hong-Xin and {Zhang}, Zhi-Yu},
        title = "{Global Dynamic Scaling Relations of H I-rich Ultra-diffuse Galaxies}",
      journal = {\apjl},
         year = 2023,
        month = apr,
       volume = {947},
       number = {1},
          eid = {L9},
        pages = {L9},
          doi = {10.3847/2041-8213/acc7a4},
archivePrefix = {arXiv},
       eprint = {2303.16232},
 primaryClass = {astro-ph.GA},
       adsurl = {https://ui.adsabs.harvard.edu/abs/2023ApJ...947L...9H}
}

@ARTICLE{rpd1,
       author = {{Jones}, Michael G. and {Sand}, David J. and {Bellazzini}, Michele and {Spekkens}, Kristine and {Karunakaran}, Ananthan and {Adams}, Elizabeth A.~K. and {Battaglia}, Giuseppina and {Beccari}, Giacomo and {Bennet}, Paul and {Cannon}, John M. and {Cresci}, Giovanni and {Crnojevi{\'c}}, Denija and {Caldwell}, Nelson and {Fuson}, Jackson and {Guhathakurta}, Puragra and {Haynes}, Martha P. and {Inoue}, John L. and {Magrini}, Laura and {Mu{\~n}oz}, Ricardo R. and {Mutlu-Pakdil}, Bur{\c{c}}in and {Seth}, Anil and {Strader}, Jay and {Toloba}, Elisa and {Zaritsky}, Dennis},
        title = "{Young, Blue, and Isolated Stellar Systems in the Virgo Cluster. II. A New Class of Stellar System}",
      journal = {\apj},
         year = 2022,
        month = aug,
       volume = {935},
       number = {1},
          eid = {51},
        pages = {51},
          doi = {10.3847/1538-4357/ac7c6c},
archivePrefix = {arXiv},
       eprint = {2205.01695},
 primaryClass = {astro-ph.GA},
       adsurl = {https://ui.adsabs.harvard.edu/abs/2022ApJ...935...51J}
}

@ARTICLE{rpd2,
       author = {{Dey}, Swapnaneel and {Jones}, Michael G. and {Sand}, David J. and {Mazziotti}, Nicolas and {Janowiecki}, Steven and {Zeimann}, Gregory R. and {Bennet}, Paul},
        title = "{Citizen Science Identification of Isolated Blue Stellar Systems in the Virgo Cluster}",
      journal = {\apj},
         year = 2025,
        month = apr,
       volume = {983},
       number = {1},
          eid = {2},
        pages = {2},
          doi = {10.3847/1538-4357/adbbd8},
archivePrefix = {arXiv},
       eprint = {2411.14526},
 primaryClass = {astro-ph.GA},
       adsurl = {https://ui.adsabs.harvard.edu/abs/2025ApJ...983....2D}
}

@ARTICLE{UDGX-1,
       author = {{Junais} and {Boissier}, S. and {Boselli}, A. and {Boquien}, M. and {Longobardi}, A. and {Roehlly}, Y. and {Amram}, P. and {Fossati}, M. and {Cuillandre}, J. -C. and {Gwyn}, S. and {Ferrarese}, L. and {C{\^o}t{\'e}}, P. and {Roediger}, J. and {Lim}, S. and {Peng}, E.~W. and {Hensler}, G. and {Trinchieri}, G. and {Koda}, J. and {Prantzos}, N.},
        title = "{A Virgo Environmental Survey Tracing Ionised Gas Emission (VESTIGE). X. Formation of a red ultra-diffuse galaxy and an almost dark galaxy during a ram-pressure stripping event}",
      journal = {\aap},
         year = 2021,
        month = jun,
       volume = {650},
          eid = {A99},
        pages = {A99},
          doi = {10.1051/0004-6361/202040185},
archivePrefix = {arXiv},
       eprint = {2104.02492},
 primaryClass = {astro-ph.GA},
       adsurl = {https://ui.adsabs.harvard.edu/abs/2021A&A...650A..99J}
}

@ARTICLE{UDGX-2,
       author = {{Jones}, Michael G. and {Sand}, David J. and {Bellazzini}, Michele and {Spekkens}, Kristine and {Cannon}, John M. and {Mutlu-Pakdil}, Bur{\c{c}}in and {Karunakaran}, Ananthan and {Beccari}, Giacomo and {Magrini}, Laura and {Cresci}, Giovanni and {Inoue}, John L. and {Fuson}, Jackson and {Adams}, Elizabeth A.~K. and {Battaglia}, Giuseppina and {Bennet}, Paul and {Crnojevi{\'c}}, Denija and {Caldwell}, Nelson and {Guhathakurta}, Puragra and {Haynes}, Martha P. and {Mu{\~n}oz}, Ricardo R. and {Seth}, Anil and {Strader}, Jay and {Toloba}, Elisa and {Zaritsky}, Dennis},
        title = "{AGC 226178 and NGVS 3543: Two Deceptive Dwarfs toward Virgo}",
      journal = {\apjl},
         year = 2022,
        month = feb,
       volume = {926},
       number = {2},
          eid = {L15},
        pages = {L15},
          doi = {10.3847/2041-8213/ac51dc},
archivePrefix = {arXiv},
       eprint = {2110.14648},
 primaryClass = {astro-ph.GA},
       adsurl = {https://ui.adsabs.harvard.edu/abs/2022ApJ...926L..15J}
}

@ARTICLE{UDGX-1correct,
       author = {{Junais} and {Boissier}, S. and {Boselli}, A. and {Ferrarese}, L. and {C{\^o}t{\'e}}, P. and {Gwyn}, S. and {Roediger}, J. and {Lim}, S. and {Peng}, E.~W. and {Cuillandre}, J. -C. and {Longobardi}, A. and {Fossati}, M. and {Hensler}, G. and {Koda}, J. and {Bautista}, J. and {Boquien}, M. and {Ma{\l}ek}, K. and {Amram}, P. and {Roehlly}, Y.},
        title = "{A Virgo Environmental Survey Tracing Ionised Gas Emission (VESTIGE). XIII. The role of ram-pressure stripping in transforming the diffuse and ultra-diffuse galaxies in the Virgo cluster}",
      journal = {\aap},
         year = 2022,
        month = nov,
       volume = {667},
          eid = {A76},
        pages = {A76},
          doi = {10.1051/0004-6361/202244237},
archivePrefix = {arXiv},
       eprint = {2208.02634},
 primaryClass = {astro-ph.GA},
       adsurl = {https://ui.adsabs.harvard.edu/abs/2022A&A...667A..76J}
}

@ARTICLE{UDGX-3,
       author = {{Sun}, Yu-Zhu and {Zhang}, Hong-Xin and {Brinks}, Elias and {Smith}, Rory and {Li}, Fujia and {Kim}, Minsu and {Oh}, Se-Heon and {Lin}, Zesen and {Kim}, Jaebeom and {Sun}, Weibin and {Li}, Tie and {C{\^o}t{\'e}}, Patrick and {Boselli}, Alessandro and {Chen}, Lijun and {Duc}, Pierre-Alain and {Paudel}, Sanjaya and {Taylor}, Matthew A. and {Wang}, Kaixiang and {Wang}, Enci and {Zhang}, Lanyue and {Zhao}, Yinghe},
        title = "{Unveiling the nature and fate of the almost-dark cloud AGC 226178 through H I mapping}",
      journal = {\aap},
         year = 2025,
        month = sep,
       volume = {701},
          eid = {A73},
        pages = {A73},
          doi = {10.1051/0004-6361/202555150},
archivePrefix = {arXiv},
       eprint = {2506.23232},
 primaryClass = {astro-ph.GA},
       adsurl = {https://ui.adsabs.harvard.edu/abs/2025A&A...701A..73S}
}

@ARTICLE{VCC1964_AA,
       author = {{di Serego Alighieri}, S. and {Gavazzi}, G. and {Giovanardi}, C. and {Giovanelli}, R. and {Grossi}, M. and {Haynes}, M.~P. and {Kent}, B.~R. and {Koopmann}, R.~A. and {Pellegrini}, S. and {Scodeggio}, M. and {Trinchieri}, G.},
        title = "{The HI content of early-type galaxies from the ALFALFA survey. I. Catalogued HI sources in the Virgo cluster}",
      journal = {\aap},
         year = 2007,
        month = nov,
       volume = {474},
       number = {3},
        pages = {851-855},
          doi = {10.1051/0004-6361:20078205},
archivePrefix = {arXiv},
       eprint = {0709.2096},
 primaryClass = {astro-ph},
       adsurl = {https://ui.adsabs.harvard.edu/abs/2007A&A...474..851D}
}

@ARTICLE{ALFALFA,
       author = {{Giovanelli}, Riccardo and {Haynes}, Martha P. and {Kent}, Brian R. and {Perillat}, Philip and {Saintonge}, Amelie and {Brosch}, Noah and {Catinella}, Barbara and {Hoffman}, G. Lyle and {Stierwalt}, Sabrina and {Spekkens}, Kristine and {Lerner}, Mikael S. and {Masters}, Karen L. and {Momjian}, Emmanuel and {Rosenberg}, Jessica L. and {Springob}, Christopher M. and {Boselli}, Alessandro and {Charmandaris}, Vassilis and {Darling}, Jeremy K. and {Davies}, Jonathan and {Garcia Lambas}, Diego and {Gavazzi}, Giuseppe and {Giovanardi}, Carlo and {Hardy}, Eduardo and {Hunt}, Leslie K. and {Iovino}, Angela and {Karachentsev}, Igor D. and {Karachentseva}, Valentina E. and {Koopmann}, Rebecca A. and {Marinoni}, Christian and {Minchin}, Robert and {Muller}, Erik and {Putman}, Mary and {Pantoja}, Carmen and {Salzer}, John J. and {Scodeggio}, Marco and {Skillman}, Evan and {Solanes}, Jose M. and {Valotto}, Carlos and {van Driel}, Wim and {van Zee}, Liese},
        title = "{The Arecibo Legacy Fast ALFA Survey. I. Science Goals, Survey Design, and Strategy}",
      journal = {\aj},
         year = 2005,
        month = dec,
       volume = {130},
       number = {6},
        pages = {2598-2612},
          doi = {10.1086/497431},
archivePrefix = {arXiv},
       eprint = {astro-ph/0508301},
 primaryClass = {astro-ph},
       adsurl = {https://ui.adsabs.harvard.edu/abs/2005AJ....130.2598G}
}

@ARTICLE{SMUDGES,
       author = {{Zaritsky}, Dennis and {Donnerstein}, Richard and {Dey}, Arjun and {Karunakaran}, Ananthan and {Kadowaki}, Jennifer and {Khim}, Donghyeon J. and {Spekkens}, Kristine and {Zhang}, Huanian},
        title = "{Systematically Measuring Ultra-diffuse Galaxies (SMUDGes). V. The Complete SMUDGes Catalog and the Nature of Ultradiffuse Galaxies}",
      journal = {\apjs},
         year = 2023,
        month = aug,
       volume = {267},
       number = {2},
          eid = {27},
        pages = {27},
          doi = {10.3847/1538-4365/acdd71},
archivePrefix = {arXiv},
       eprint = {2306.01524},
 primaryClass = {astro-ph.GA},
       adsurl = {https://ui.adsabs.harvard.edu/abs/2023ApJS..267...27Z}
}

@ARTICLE{SourceExtraction,
       author = {{Taylor}, Rhys},
        title = "{Quantifying the completeness and reliability of visual source extraction: An examination of eight thousand data cubes by eye}",
      journal = {\aap},
         year = 2025,
        month = apr,
       volume = {696},
          eid = {A113},
        pages = {A113},
          doi = {10.1051/0004-6361/202451606},
archivePrefix = {arXiv},
       eprint = {2503.07430},
 primaryClass = {astro-ph.IM},
       adsurl = {https://ui.adsabs.harvard.edu/abs/2025A&A...696A.113T}
}

@ARTICLE{goldmine,
       author = {{Gavazzi}, G. and {Boselli}, A. and {Donati}, A. and {Franzetti}, P. and {Scodeggio}, M.},
        title = "{Introducing GOLDMine: A new galaxy database on the WEB}",
      journal = {\aap},
         year = 2003,
        month = mar,
       volume = {400},
        pages = {451-455},
          doi = {10.1051/0004-6361:20030026},
archivePrefix = {arXiv},
       eprint = {astro-ph/0212257},
 primaryClass = {astro-ph},
       adsurl = {https://ui.adsabs.harvard.edu/abs/2003A&A...400..451G}
}

@ARTICLE{theVCC,
       author = {{Binggeli}, B. and {Sandage}, A. and {Tammann}, G.~A.},
        title = "{Studies of the Virgo cluster. II. A catalog of 2096 galaxies in the Virgo cluster area.}",
      journal = {\aj},
         year = 1985,
        month = sep,
       volume = {90},
        pages = {1681-1758},
          doi = {10.1086/113874},
       adsurl = {https://ui.adsabs.harvard.edu/abs/1985AJ.....90.1681B}
}

@ARTICLE{LS,
       author = {{Dey}, Arjun and {Schlegel}, David J. and {Lang}, Dustin and {Blum}, Robert and {Burleigh}, Kaylan and {Fan}, Xiaohui and {Findlay}, Joseph R. and {Finkbeiner}, Doug and {Herrera}, David and {Juneau}, St{\'e}phanie and {Landriau}, Martin and {Levi}, Michael and {McGreer}, Ian and {Meisner}, Aaron and {Myers}, Adam D. and {Moustakas}, John and {Nugent}, Peter and {Patej}, Anna and {Schlafly}, Edward F. and {Walker}, Alistair R. and {Valdes}, Francisco and {Weaver}, Benjamin A. and {Y{\`e}che}, Christophe and {Zou}, Hu and {Zhou}, Xu and {Abareshi}, Behzad and {Abbott}, T.~M.~C. and {Abolfathi}, Bela and {Aguilera}, C. and {Alam}, Shadab and {Allen}, Lori and {Alvarez}, A. and {Annis}, James and {Ansarinejad}, Behzad and {Aubert}, Marie and {Beechert}, Jacqueline and {Bell}, Eric F. and {BenZvi}, Segev Y. and {Beutler}, Florian and {Bielby}, Richard M. and {Bolton}, Adam S. and {Brice{\~n}o}, C{\'e}sar and {Buckley-Geer}, Elizabeth J. and {Butler}, Karen and {Calamida}, Annalisa and {Carlberg}, Raymond G. and {Carter}, Paul and {Casas}, Ricard and {Castander}, Francisco J. and {Choi}, Yumi and {Comparat}, Johan and {Cukanovaite}, Elena and {Delubac}, Timoth{\'e}e and {DeVries}, Kaitlin and {Dey}, Sharmila and {Dhungana}, Govinda and {Dickinson}, Mark and {Ding}, Zhejie and {Donaldson}, John B. and {Duan}, Yutong and {Duckworth}, Christopher J. and {Eftekharzadeh}, Sarah and {Eisenstein}, Daniel J. and {Etourneau}, Thomas and {Fagrelius}, Parker A. and {Farihi}, Jay and {Fitzpatrick}, Mike and {Font-Ribera}, Andreu and {Fulmer}, Leah and {G{\"a}nsicke}, Boris T. and {Gaztanaga}, Enrique and {George}, Koshy and {Gerdes}, David W. and {Gontcho}, Satya Gontcho A. and {Gorgoni}, Claudio and {Green}, Gregory and {Guy}, Julien and {Harmer}, Diane and {Hernandez}, M. and {Honscheid}, Klaus and {Huang}, Lijuan Wendy and {James}, David J. and {Jannuzi}, Buell T. and {Jiang}, Linhua and {Joyce}, Richard and {Karcher}, Armin and {Karkar}, Sonia and {Kehoe}, Robert and {Kneib}, Jean-Paul and {Kueter-Young}, Andrea and {Lan}, Ting-Wen and {Lauer}, Tod R. and {Le Guillou}, Laurent and {Le Van Suu}, Auguste and {Lee}, Jae Hyeon and {Lesser}, Michael and {Perreault Levasseur}, Laurence and {Li}, Ting S. and {Mann}, Justin L. and {Marshall}, Robert and {Mart{\'\i}nez-V{\'a}zquez}, C.~E. and {Martini}, Paul and {du Mas des Bourboux}, H{\'e}lion and {McManus}, Sean and {Meier}, Tobias Gabriel and {M{\'e}nard}, Brice and {Metcalfe}, Nigel and {Mu{\~n}oz-Guti{\'e}rrez}, Andrea and {Najita}, Joan and {Napier}, Kevin and {Narayan}, Gautham and {Newman}, Jeffrey A. and {Nie}, Jundan and {Nord}, Brian and {Norman}, Dara J. and {Olsen}, Knut A.~G. and {Paat}, Anthony and {Palanque-Delabrouille}, Nathalie and {Peng}, Xiyan and {Poppett}, Claire L. and {Poremba}, Megan R. and {Prakash}, Abhishek and {Rabinowitz}, David and {Raichoor}, Anand and {Rezaie}, Mehdi and {Robertson}, A.~N. and {Roe}, Natalie A. and {Ross}, Ashley J. and {Ross}, Nicholas P. and {Rudnick}, Gregory and {Safonova}, Sasha and {Saha}, Abhijit and {S{\'a}nchez}, F. Javier and {Savary}, Elodie and {Schweiker}, Heidi and {Scott}, Adam and {Seo}, Hee-Jong and {Shan}, Huanyuan and {Silva}, David R. and {Slepian}, Zachary and {Soto}, Christian and {Sprayberry}, David and {Staten}, Ryan and {Stillman}, Coley M. and {Stupak}, Robert J. and {Summers}, David L. and {Sien Tie}, Suk and {Tirado}, H. and {Vargas-Maga{\~n}a}, Mariana and {Vivas}, A. Katherina and {Wechsler}, Risa H. and {Williams}, Doug and {Yang}, Jinyi and {Yang}, Qian and {Yapici}, Tolga and {Zaritsky}, Dennis and {Zenteno}, A. and {Zhang}, Kai and {Zhang}, Tianmeng and {Zhou}, Rongpu and {Zhou}, Zhimin},
        title = "{Overview of the DESI Legacy Imaging Surveys}",
      journal = {\aj},
         year = 2019,
        month = may,
       volume = {157},
       number = {5},
          eid = {168},
        pages = {168},
          doi = {10.3847/1538-3881/ab089d},
archivePrefix = {arXiv},
       eprint = {1804.08657},
 primaryClass = {astro-ph.IM},
       adsurl = {https://ui.adsabs.harvard.edu/abs/2019AJ....157..168D}
}

@ARTICLE{LSdepth,
       author = {{Mart{\'\i}nez-Delgado}, David and {Cooper}, Andrew P. and {Rom{\'a}n}, Javier and {Pillepich}, Annalisa and {Erkal}, Denis and {Pearson}, Sarah and {Moustakas}, John and {Laporte}, Chervin F.~P. and {Laine}, Seppo and {Akhlaghi}, Mohammad and {Lang}, Dustin and {Makarov}, Dmitry and {Borlaff}, Alejandro S. and {Donatiello}, Giuseppe and {Pearson}, William J. and {Mir{\'o}-Carretero}, Juan and {Cuillandre}, Jean-Charles and {Dom{\'\i}nguez}, Helena and {Roca-F{\`a}brega}, Santi and {Frenk}, Carlos S. and {Schmidt}, Judy and {G{\'o}mez-Flechoso}, Mar{\'\i}a A. and {Guzman}, Rafael and {Libeskind}, Noam I. and {Dey}, Arjun and {Weaver}, Benjamin A. and {Schlegel}, David and {Myers}, Adam D. and {Valdes}, Frank G.},
        title = "{Hidden depths in the local Universe: The Stellar Stream Legacy Survey}",
      journal = {\aap},
         year = 2023,
        month = mar,
       volume = {671},
          eid = {A141},
        pages = {A141},
          doi = {10.1051/0004-6361/202245011},
archivePrefix = {arXiv},
       eprint = {2104.06071},
 primaryClass = {astro-ph.GA},
       adsurl = {https://ui.adsabs.harvard.edu/abs/2023A&A...671A.141M}
}

@ARTICLE{AASDSS,
       author = {{Durbala}, Adriana and {Finn}, Rose A. and {Crone Odekon}, Mary and {Haynes}, Martha P. and {Koopmann}, Rebecca A. and {O'Donoghue}, Aileen A.},
        title = "{The ALFALFA-SDSS Galaxy Catalog}",
      journal = {\aj},
         year = 2020,
        month = dec,
       volume = {160},
       number = {6},
          eid = {271},
        pages = {271},
          doi = {10.3847/1538-3881/abc018},
archivePrefix = {arXiv},
       eprint = {2011.02588},
 primaryClass = {astro-ph.GA},
       adsurl = {https://ui.adsabs.harvard.edu/abs/2020AJ....160..271D}
}

@ARTICLE{kinky,
       author = {{Taylor}, R. and {Davies}, J.~I. and {J{\'a}chym}, P. and {Keenan}, O. and {Minchin}, R.~F. and {Palou{\v{s}}}, J. and {Smith}, R. and {W{\"u}nsch}, R.},
        title = "{Kinematic clues to the origins of starless H I clouds: dark galaxies or tidal debris?}",
      journal = {\mnras},
         year = 2017,
        month = may,
       volume = {467},
       number = {3},
        pages = {3648-3661},
          doi = {10.1093/mnras/stx187},
archivePrefix = {arXiv},
       eprint = {1701.05361},
 primaryClass = {astro-ph.GA},
       adsurl = {https://ui.adsabs.harvard.edu/abs/2017MNRAS.467.3648T}
}

@ARTICLE{BTFR,
       author = {{McGaugh}, S.~S. and {Schombert}, J.~M. and {Bothun}, G.~D. and {de Blok}, W.~J.~G.},
        title = "{The Baryonic Tully-Fisher Relation}",
      journal = {\apjl},
         year = 2000,
        month = apr,
       volume = {533},
       number = {2},
        pages = {L99-L102},
          doi = {10.1086/312628},
archivePrefix = {arXiv},
       eprint = {astro-ph/0003001},
 primaryClass = {astro-ph},
       adsurl = {https://ui.adsabs.harvard.edu/abs/2000ApJ...533L..99M}
}

@ARTICLE{distanceproblems,
       author = {{Beasley}, M.~A. and {Fahrion}, K. and {Guerra Arencibia}, S. and {Gvozdenko}, A. and {Montes}, M.},
        title = "{A new way to measure the distance to NGC1052-DF2}",
      journal = {\aap},
         year = 2025,
        month = may,
       volume = {697},
          eid = {A144},
        pages = {A144},
          doi = {10.1051/0004-6361/202452446},
archivePrefix = {arXiv},
       eprint = {2503.03403},
 primaryClass = {astro-ph.GA},
       adsurl = {https://ui.adsabs.harvard.edu/abs/2025A&A...697A.144B}
}

@ARTICLE{blobsim,
       author = {{Calura}, Francesco and {Bellazzini}, Michele and {D'Ercole}, Annibale},
        title = "{Hydrodynamic simulations of an isolated star-forming gas cloud in the Virgo cluster}",
      journal = {\mnras},
         year = 2020,
        month = dec,
       volume = {499},
       number = {4},
        pages = {5873-5890},
          doi = {10.1093/mnras/staa3133},
archivePrefix = {arXiv},
       eprint = {2010.09758},
 primaryClass = {astro-ph.GA},
       adsurl = {https://ui.adsabs.harvard.edu/abs/2020MNRAS.499.5873C}
}

@ARTICLE{VCC2037,
       author = {{Karachentsev}, Igor. D. and {Tully}, R. Brent and {Wu}, Po-Feng and {Shaya}, Edward J. and {Dolphin}, Andrew E.},
        title = "{Infall of Nearby Galaxies into the Virgo Cluster as Traced with Hubble Space Telescope}",
      journal = {\apj},
         year = 2014,
        month = feb,
       volume = {782},
       number = {1},
          eid = {4},
        pages = {4},
          doi = {10.1088/0004-637X/782/1/4},
archivePrefix = {arXiv},
       eprint = {1312.6769},
 primaryClass = {astro-ph.GA},
       adsurl = {https://ui.adsabs.harvard.edu/abs/2014ApJ...782....4K}
}

@ARTICLE{FRELLED5,
       author = {{Taylor}, R.},
        title = "{FRELLED Reloaded: Multiple techniques for astronomical data visualisation in Blender}",
      journal = {Astronomy and Computing},
         year = 2025,
        month = apr,
       volume = {51},
          eid = {100927},
        pages = {100927},
          doi = {10.1016/j.ascom.2024.100927},
archivePrefix = {arXiv},
       eprint = {2501.02919},
 primaryClass = {astro-ph.IM},
       adsurl = {https://ui.adsabs.harvard.edu/abs/2025A&C....5100927T}
}

@PHDTHESIS{mythesis,
       author = {{Taylor}, Rhys},
        title = "{Virgo Cluster through the AGES}",
       school = {Cardiff University, UK},
         year = 2010,
        month = jan,
       adsurl = {https://ui.adsabs.harvard.edu/abs/2010PhDT.......208T}
}

@ARTICLE{BTFRSample,
       author = {{McGaugh}, Stacy S.},
        title = "{The Baryonic Tully-Fisher Relation of Gas-rich Galaxies as a Test of {\ensuremath{\Lambda}}CDM and MOND}",
      journal = {\aj},
         year = 2012,
        month = feb,
       volume = {143},
       number = {2},
          eid = {40},
        pages = {40},
          doi = {10.1088/0004-6256/143/2/40},
archivePrefix = {arXiv},
       eprint = {1107.2934},
 primaryClass = {astro-ph.CO},
       adsurl = {https://ui.adsabs.harvard.edu/abs/2012AJ....143...40M}
}

@ARTICLE{bell,
       author = {{Bell}, Eric F. and {McIntosh}, Daniel H. and {Katz}, Neal and {Weinberg}, Martin D.},
        title = "{The Optical and Near-Infrared Properties of Galaxies. I. Luminosity and Stellar Mass Functions}",
      journal = {\apjs},
         year = 2003,
        month = dec,
       volume = {149},
       number = {2},
        pages = {289-312},
          doi = {10.1086/378847},
archivePrefix = {arXiv},
       eprint = {astro-ph/0302543},
 primaryClass = {astro-ph},
       adsurl = {https://ui.adsabs.harvard.edu/abs/2003ApJS..149..289B}
}

@ARTICLE{taylormass,
       author = {{Taylor}, Edward N. and {Hopkins}, Andrew M. and {Baldry}, Ivan K. and {Brown}, Michael J.~I. and {Driver}, Simon P. and {Kelvin}, Lee S. and {Hill}, David T. and {Robotham}, Aaron S.~G. and {Bland-Hawthorn}, Joss and {Jones}, D.~H. and {Sharp}, R.~G. and {Thomas}, Daniel and {Liske}, Jochen and {Loveday}, Jon and {Norberg}, Peder and {Peacock}, J.~A. and {Bamford}, Steven P. and {Brough}, Sarah and {Colless}, Matthew and {Cameron}, Ewan and {Conselice}, Christopher J. and {Croom}, Scott M. and {Frenk}, C.~S. and {Gunawardhana}, Madusha and {Kuijken}, Konrad and {Nichol}, R.~C. and {Parkinson}, H.~R. and {Phillipps}, S. and {Pimbblet}, K.~A. and {Popescu}, C.~C. and {Prescott}, Matthew and {Sutherland}, W.~J. and {Tuffs}, R.~J. and {van Kampen}, Eelco and {Wijesinghe}, D.},
        title = "{Galaxy And Mass Assembly (GAMA): stellar mass estimates}",
      journal = {\mnras},
         year = 2011,
        month = dec,
       volume = {418},
       number = {3},
        pages = {1587-1620},
          doi = {10.1111/j.1365-2966.2011.19536.x},
archivePrefix = {arXiv},
       eprint = {1108.0635},
 primaryClass = {astro-ph.CO},
       adsurl = {https://ui.adsabs.harvard.edu/abs/2011MNRAS.418.1587T}
}

@ARTICLE{T15,
       author = {{Taylor}, R.},
        title = "{FRELLED: A realtime volumetric data viewer for astronomers}",
      journal = {Astronomy and Computing},
         year = 2015,
        month = nov,
       volume = {13},
        pages = {67-79},
          doi = {10.1016/j.ascom.2015.10.002},
archivePrefix = {arXiv},
       eprint = {1510.03589},
 primaryClass = {astro-ph.IM},
       adsurl = {https://ui.adsabs.harvard.edu/abs/2015A&C....13...67T}
}

@ARTICLE{agesvc1,
       author = {{Taylor}, R. and {Davies}, J.~I. and {Auld}, R. and {Minchin}, R.~F.},
        title = "{The Arecibo Galaxy Environment Survey - V. The Virgo cluster (I)}",
      journal = {\mnras},
         year = 2012,
        month = jun,
       volume = {423},
       number = {1},
        pages = {787-810},
          doi = {10.1111/j.1365-2966.2012.20914.x},
archivePrefix = {arXiv},
       eprint = {1203.3094},
 primaryClass = {astro-ph.GA},
       adsurl = {https://ui.adsabs.harvard.edu/abs/2012MNRAS.423..787T}
}

@ARTICLE{AGESM33,
       author = {{Keenan}, O.~C. and {Davies}, J.~I. and {Taylor}, R. and {Minchin}, R.~F.},
        title = "{The Arecibo Galaxy Environment Survey - X. The structure of halo gas around M33}",
      journal = {\mnras},
         year = 2016,
        month = feb,
       volume = {456},
       number = {1},
        pages = {951-960},
          doi = {10.1093/mnras/stv2684},
archivePrefix = {arXiv},
       eprint = {1511.02710},
 primaryClass = {astro-ph.GA},
       adsurl = {https://ui.adsabs.harvard.edu/abs/2016MNRAS.456..951K}
}

@ARTICLE{AGESLeo,
       author = {{Taylor}, Rhys and {K{\"o}ppen}, Joachim and {J{\'a}chym}, Pavel and {Minchin}, Robert and {Palou{\v{s}}}, Jan and {Rosenberg}, Jessica L. and {Schneider}, Stephen and {W{\"u}nsch}, Richard and {Deshev}, Boris},
        title = "{The Arecibo Galaxy Environment Survey. XII. Optically Dark H I Clouds in the Leo I Group}",
      journal = {\aj},
         year = 2022,
        month = dec,
       volume = {164},
       number = {6},
          eid = {233},
        pages = {233},
          doi = {10.3847/1538-3881/ac96e8},
archivePrefix = {arXiv},
       eprint = {2209.10994},
 primaryClass = {astro-ph.GA},
       adsurl = {https://ui.adsabs.harvard.edu/abs/2022AJ....164..233T}
}

@INPROCEEDINGS{miriad,
       author = {{Sault}, R.~J. and {Teuben}, P.~J. and {Wright}, M.~C.~H.},
        title = "{A Retrospective View of MIRIAD}",
    booktitle = {Astronomical Data Analysis Software and Systems IV},
         year = 1995,
       editor = {{Shaw}, R.~A. and {Payne}, H.~E. and {Hayes}, J.~J.~E.},
       series = {Astronomical Society of the Pacific Conference Series},
       volume = {77},
        month = jan,
        pages = {433},
          doi = {10.48550/arXiv.astro-ph/0612759},
archivePrefix = {arXiv},
       eprint = {astro-ph/0612759},
 primaryClass = {astro-ph},
       adsurl = {https://ui.adsabs.harvard.edu/abs/1995ASPC...77..433S}
}

@ARTICLE{fading,
       author = {{Taylor}, Rhys and {K{\"o}ppen}, Joachim and {J{\'a}chym}, Pavel and {Minchin}, Robert and {Palou{\v{s}}}, Jan and {W{\"u}nsch}, Richard},
        title = "{Faint and Fading Tails: The Fate of Stripped H I Gas in Virgo Cluster Galaxies}",
      journal = {\aj},
         year = 2020,
        month = may,
       volume = {159},
       number = {5},
          eid = {218},
        pages = {218},
          doi = {10.3847/1538-3881/ab6988},
archivePrefix = {arXiv},
       eprint = {2001.03385},
 primaryClass = {astro-ph.GA},
       adsurl = {https://ui.adsabs.harvard.edu/abs/2020AJ....159..218T}
}

@ARTICLE{agesvii,
       author = {{Taylor}, R. and {Minchin}, R.~F. and {Herbst}, H. and {Davies}, J.~I. and {Rodriguez}, R. and {Vazquez}, C.},
        title = "{The Arecibo Galaxy Environment Survey - VII. A dense filament with extremely long H I streams}",
      journal = {\mnras},
         year = 2014,
        month = sep,
       volume = {443},
       number = {3},
        pages = {2634-2649},
          doi = {10.1093/mnras/stu1305},
archivePrefix = {arXiv},
       eprint = {1407.0016},
 primaryClass = {astro-ph.GA},
       adsurl = {https://ui.adsabs.harvard.edu/abs/2014MNRAS.443.2634T}
}

@ARTICLE{isolated,
       author = {{Minchin}, R.~F. and {Momjian}, E. and {Auld}, R. and {Davies}, J.~I. and {Valls-Gabaud}, D. and {Karachentsev}, I.~D. and {Henning}, P.~A. and {O'Neil}, K.~L. and {Schneider}, S. and {Smith}, M.~W.~L. and {Stage}, M.~D. and {Taylor}, R. and {van Driel}, W.},
        title = "{The Arecibo Galaxy Environment Survey. III. Observations Toward the Galaxy Pair NGC 7332/7339 and the Isolated Galaxy NGC 1156}",
      journal = {\aj},
         year = 2010,
        month = oct,
       volume = {140},
       number = {4},
        pages = {1093-1118},
          doi = {10.1088/0004-6256/140/4/1093},
archivePrefix = {arXiv},
       eprint = {1009.4438},
 primaryClass = {astro-ph.CO},
       adsurl = {https://ui.adsabs.harvard.edu/abs/2010AJ....140.1093M}
}

@ARTICLE{ages,
       author = {{Auld}, R. and {Minchin}, R.~F. and {Davies}, J.~I. and {Catinella}, B. and {van Driel}, W. and {Henning}, P.~A. and {Linder}, S. and {Momjian}, E. and {Muller}, E. and {O'Neil}, K. and {Sabatini}, S. and {Schneider}, S. and {Bothun}, G. and {Cortese}, L. and {Disney}, M. and {Hoffman}, G.~L. and {Putman}, M. and {Rosenberg}, J.~L. and {Baes}, M. and {de Blok}, W.~J.~G. and {Boselli}, A. and {Brinks}, E. and {Brosch}, N. and {Irwin}, J. and {Karachentsev}, I.~D. and {Kilborn}, V.~A. and {Koribalski}, B. and {Spekkens}, K.},
        title = "{The Arecibo Galaxy Environment Survey: precursor observations of the NGC 628 group}",
      journal = {\mnras},
         year = 2006,
        month = oct,
       volume = {371},
       number = {4},
        pages = {1617-1640},
          doi = {10.1111/j.1365-2966.2006.10761.x},
archivePrefix = {arXiv},
       eprint = {astro-ph/0607452},
 primaryClass = {astro-ph},
       adsurl = {https://ui.adsabs.harvard.edu/abs/2006MNRAS.371.1617A}
}

@ARTICLE{saint,
       author = {{Saintonge}, Am{\'e}lie},
        title = "{The Arecibo Legacy Fast ALFA Survey. IV. Strategies for Signal Identification and Survey Catalog Reliability}",
      journal = {\aj},
         year = 2007,
        month = may,
       volume = {133},
       number = {5},
        pages = {2087-2096},
          doi = {10.1086/513515},
archivePrefix = {arXiv},
       eprint = {astro-ph/0702178},
 primaryClass = {astro-ph},
       adsurl = {https://ui.adsabs.harvard.edu/abs/2007AJ....133.2087S}
}

@ARTICLE{agesvc2,
       author = {{Taylor}, R. and {Davies}, J.~I. and {Auld}, R. and {Minchin}, R.~F. and {Smith}, R.},
        title = "{The Arecibo Galaxy Environment Survey - VI. The Virgo cluster (II)}",
      journal = {\mnras},
         year = 2013,
        month = jan,
       volume = {428},
       number = {1},
        pages = {459-469},
          doi = {10.1093/mnras/sts042},
archivePrefix = {arXiv},
       eprint = {1209.4338},
 primaryClass = {astro-ph.GA},
       adsurl = {https://ui.adsabs.harvard.edu/abs/2013MNRAS.428..459T}
}

@ARTICLE{waves,
       author = {{Minchin}, Robert F. and {Taylor}, Rhys and {K{\"o}ppen}, Joachim and {Davies}, Jonathan I. and {van Driel}, Wim and {Keenan}, Olivia},
        title = "{The Widefield Arecibo Virgo Extragalactic Survey. I. New Structures in the ALFALFA Virgo 7 Cloud Complex and an Extended Tail on NGC 4522}",
      journal = {\aj},
         year = 2019,
        month = sep,
       volume = {158},
       number = {3},
          eid = {121},
        pages = {121},
          doi = {10.3847/1538-3881/ab303e},
archivePrefix = {arXiv},
       eprint = {1907.07217},
 primaryClass = {astro-ph.GA},
       adsurl = {https://ui.adsabs.harvard.edu/abs/2019AJ....158..121M}
}

@ARTICLE{boris,
       author = {{Deshev}, Boris and {Taylor}, Rhys and {Minchin}, Robert and {Scott}, Tom C. and {Brinks}, Elias},
        title = "{The Arecibo Galaxy Environment Survey (AGES). XI. The expanded Abell 1367 field: Data catalogue and H I census over the surveyed volume}",
      journal = {\aap},
         year = 2022,
        month = sep,
       volume = {665},
          eid = {A155},
        pages = {A155},
          doi = {10.1051/0004-6361/202243103},
archivePrefix = {arXiv},
       eprint = {2206.13533},
 primaryClass = {astro-ph.GA},
       adsurl = {https://ui.adsabs.harvard.edu/abs/2022A&A...665A.155D}
}

@ARTICLE{agesiv,
       author = {{Davies}, J.~I. and {Auld}, R. and {Burns}, L. and {Minchin}, R. and {Momjian}, E. and {Schneider}, S. and {Smith}, M. and {Taylor}, R. and {van Driel}, W.},
        title = "{The Arecibo Galaxy Environment Survey - IV. The NGC 7448 region and the H I mass function}",
      journal = {\mnras},
         year = 2011,
        month = aug,
       volume = {415},
       number = {2},
        pages = {1883-1894},
          doi = {10.1111/j.1365-2966.2011.18833.x},
       adsurl = {https://ui.adsabs.harvard.edu/abs/2011MNRAS.415.1883D}
}

@ARTICLE{AA100,
       author = {{Haynes}, Martha P. and {Giovanelli}, Riccardo and {Kent}, Brian R. and {Adams}, Elizabeth A.~K. and {Balonek}, Thomas J. and {Craig}, David W. and {Fertig}, Derek and {Finn}, Rose and {Giovanardi}, Carlo and {Hallenbeck}, Gregory and {Hess}, Kelley M. and {Hoffman}, G. Lyle and {Huang}, Shan and {Jones}, Michael G. and {Koopmann}, Rebecca A. and {Kornreich}, David A. and {Leisman}, Lukas and {Miller}, Jeffrey and {Moorman}, Crystal and {O'Connor}, Jessica and {O'Donoghue}, Aileen and {Papastergis}, Emmanouil and {Troischt}, Parker and {Stark}, David and {Xiao}, Li},
        title = "{The Arecibo Legacy Fast ALFA Survey: The ALFALFA Extragalactic H I Source Catalog}",
      journal = {\apj},
         year = 2018,
        month = jul,
       volume = {861},
       number = {1},
          eid = {49},
        pages = {49},
          doi = {10.3847/1538-4357/aac956},
archivePrefix = {arXiv},
       eprint = {1805.11499},
 primaryClass = {astro-ph.GA},
       adsurl = {https://ui.adsabs.harvard.edu/abs/2018ApJ...861...49H}
}

@ARTICLE{SOFIA2,
       author = {{Westmeier}, T. and {Kitaeff}, S. and {Pallot}, D. and {Serra}, P. and {van der Hulst}, J.~M. and {Jurek}, R.~J. and {Elagali}, A. and {For}, B. -Q. and {Kleiner}, D. and {Koribalski}, B.~S. and {Lee-Waddell}, K. and {Mould}, J.~R. and {Reynolds}, T.~N. and {Rhee}, J. and {Staveley-Smith}, L.},
        title = "{SOFIA 2 - an automated, parallel H I source finding pipeline for the WALLABY survey}",
      journal = {\mnras},
         year = 2021,
        month = sep,
       volume = {506},
       number = {3},
        pages = {3962-3976},
          doi = {10.1093/mnras/stab1881},
archivePrefix = {arXiv},
       eprint = {2106.15789},
 primaryClass = {astro-ph.IM},
       adsurl = {https://ui.adsabs.harvard.edu/abs/2021MNRAS.506.3962W}
}

@ARTICLE{hogg,
       author = {{Hogg}, David W. and {Turner}, Edwin L.},
        title = "{A Maximum Likelihood Method to Improve Faint-Source Flux and Color Estimates}",
      journal = {\pasp},
         year = 1998,
        month = jun,
       volume = {110},
       number = {748},
        pages = {727-731},
          doi = {10.1086/316173},
archivePrefix = {arXiv},
       eprint = {astro-ph/9711154},
 primaryClass = {astro-ph},
       adsurl = {https://ui.adsabs.harvard.edu/abs/1998PASP..110..727H}
}

@ARTICLE{solanes,
       author = {{Solanes}, Jose M. and {Giovanelli}, Riccardo and {Haynes}, Martha P.},
        title = "{The H i Content of Spirals. I. Field Galaxy H i Mass Functions and H i Mass--Optical Size Regressions}",
      journal = {\apj},
         year = 1996,
        month = apr,
       volume = {461},
        pages = {609},
          doi = {10.1086/177089},
archivePrefix = {arXiv},
       eprint = {astro-ph/9511003},
 primaryClass = {astro-ph},
       adsurl = {https://ui.adsabs.harvard.edu/abs/1996ApJ...461..609S}
}

@ARTICLE{silkUDGs,
       author = {{Silk}, Joseph},
        title = "{Ultra-diffuse galaxies without dark matter}",
      journal = {\mnras},
         year = 2019,
        month = sep,
       volume = {488},
       number = {1},
        pages = {L24-L28},
          doi = {10.1093/mnrasl/slz090},
archivePrefix = {arXiv},
       eprint = {1905.13235},
 primaryClass = {astro-ph.GA},
       adsurl = {https://ui.adsabs.harvard.edu/abs/2019MNRAS.488L..24S}
}

@ARTICLE{UDGcollisions,
       author = {{Moreno}, Jorge and {Danieli}, Shany and {Bullock}, James S. and {Feldmann}, Robert and {Hopkins}, Philip F. and {{\c{c}}atmabacak}, Onur and {Gurvich}, Alexander and {Lazar}, Alexandres and {Klein}, Courtney and {Hummels}, Cameron B. and {Hafen}, Zachary and {Mercado}, Francisco J. and {Yu}, Sijie and {Jiang}, Fangzhou and {Wheeler}, Coral and {Wetzel}, Andrew and {Angl{\'e}s-Alc{\'a}zar}, Daniel and {Boylan-Kolchin}, Michael and {Quataert}, Eliot and {Faucher-Gigu{\`e}re}, Claude-Andr{\'e} and {Kere{\v{s}}}, Du{\v{s}}an},
        title = "{Galaxies lacking dark matter produced by close encounters in a cosmological simulation}",
      journal = {Nature Astronomy},
         year = 2022,
        month = apr,
       volume = {6},
        pages = {496-502},
          doi = {10.1038/s41550-021-01598-4},
archivePrefix = {arXiv},
       eprint = {2202.05836},
 primaryClass = {astro-ph.GA},
       adsurl = {https://ui.adsabs.harvard.edu/abs/2022NatAs...6..496M}
}

@ARTICLE{pina2020,
       author = {{Mancera Pi{\~n}a}, Pavel E. and {Fraternali}, Filippo and {Oman}, Kyle A. and {Adams}, Elizabeth A.~K. and {Bacchini}, Cecilia and {Marasco}, Antonino and {Oosterloo}, Tom and {Pezzulli}, Gabriele and {Posti}, Lorenzo and {Leisman}, Lukas and {Cannon}, John M. and {di Teodoro}, Enrico M. and {Gault}, Lexi and {Haynes}, Martha P. and {Reiter}, Kameron and {Rhode}, Katherine L. and {Salzer}, John J. and {Smith}, Nicholas J.},
        title = "{Robust H I kinematics of gas-rich ultra-diffuse galaxies: hints of a weak-feedback formation scenario}",
      journal = {\mnras},
         year = 2020,
        month = jul,
       volume = {495},
       number = {4},
        pages = {3636-3655},
          doi = {10.1093/mnras/staa1256},
archivePrefix = {arXiv},
       eprint = {2004.14392},
 primaryClass = {astro-ph.GA},
       adsurl = {https://ui.adsabs.harvard.edu/abs/2020MNRAS.495.3636M}
}

@ARTICLE{allAAUDGs,
       author = {{Janowiecki}, Steven and {Jones}, Michael G. and {Leisman}, Lukas and {Webb}, Andrew},
        title = "{The environment of H I-bearing ultra-diffuse galaxies in the ALFALFA survey}",
      journal = {\mnras},
         year = 2019,
        month = nov,
       volume = {490},
       number = {1},
        pages = {566-577},
          doi = {10.1093/mnras/stz1868},
archivePrefix = {arXiv},
       eprint = {1906.11543},
 primaryClass = {astro-ph.GA},
       adsurl = {https://ui.adsabs.harvard.edu/abs/2019MNRAS.490..566J}
}

@ARTICLE{moreHUDs,
       author = {{Spekkens}, Kristine and {Karunakaran}, Ananthan},
        title = "{Atomic Gas in Blue Ultra Diffuse Galaxies around Hickson Compact Groups}",
      journal = {\apj},
         year = 2018,
        month = mar,
       volume = {855},
       number = {1},
          eid = {28},
        pages = {28},
          doi = {10.3847/1538-4357/aa94be},
archivePrefix = {arXiv},
       eprint = {1710.06557},
 primaryClass = {astro-ph.GA},
       adsurl = {https://ui.adsabs.harvard.edu/abs/2018ApJ...855...28S}
}

@ARTICLE{onemoreHUD,
       author = {{Shi}, Dong Dong and {Zheng}, Xian Zhong and {Zhao}, Hai Bin and {Pan}, Zhi Zheng and {Li}, Bin and {Zou}, Hu and {Zhou}, Xu and {Guo}, KeXin and {An}, Fang Xia and {Li}, Yu Bin},
        title = "{Deep Imaging of the HCG 95 Field. I. Ultra-diffuse Galaxies}",
      journal = {\apj},
         year = 2017,
        month = sep,
       volume = {846},
       number = {1},
          eid = {26},
        pages = {26},
          doi = {10.3847/1538-4357/aa8327},
archivePrefix = {arXiv},
       eprint = {1708.00013},
 primaryClass = {astro-ph.GA},
       adsurl = {https://ui.adsabs.harvard.edu/abs/2017ApJ...846...26S}
}

@ARTICLE{afewmorehuds,
       author = {{Karunakaran}, Ananthan and {Spekkens}, Kristine and {Zaritsky}, Dennis and {Donnerstein}, Richard L. and {Kadowaki}, Jennifer and {Dey}, Arjun},
        title = "{Systematically Measuring Ultradiffuse Galaxies in H I: Results from the Pilot Survey}",
      journal = {\apj},
         year = 2020,
        month = oct,
       volume = {902},
       number = {1},
          eid = {39},
        pages = {39},
          doi = {10.3847/1538-4357/abb464},
archivePrefix = {arXiv},
       eprint = {2005.14202},
 primaryClass = {astro-ph.GA},
       adsurl = {https://ui.adsabs.harvard.edu/abs/2020ApJ...902...39K}
}

@ARTICLE{fieldUPSB,
       author = {{Sandoval Ascencio}, Loraine and {Cooper}, M.~C. and {Zaritsky}, Dennis and {Donnerstein}, Richard and {Khim}, Donghyeon J. and {Baxter}, Devontae C.},
        title = "{Caught in the Act of Quenching? {\textendash} A Population of Post-Starburst Ultra-Diffuse Galaxies}",
      journal = {The Open Journal of Astrophysics},
         year = 2025,
        month = aug,
       volume = {8},
          eid = {110},
        pages = {110},
          doi = {10.33232/001c.142946},
archivePrefix = {arXiv},
       eprint = {2502.00117},
 primaryClass = {astro-ph.GA},
       adsurl = {https://ui.adsabs.harvard.edu/abs/2025OJAp....8E.110S}
}

@ARTICLE{wronginc,
       author = {{Oman}, Kyle A. and {Navarro}, Julio F. and {Sales}, Laura V. and {Fattahi}, Azadeh and {Frenk}, Carlos S. and {Sawala}, Till and {Schaller}, Matthieu and {White}, Simon D.~M.},
        title = "{Missing dark matter in dwarf galaxies?}",
      journal = {\mnras},
         year = 2016,
        month = aug,
       volume = {460},
       number = {4},
        pages = {3610-3623},
          doi = {10.1093/mnras/stw1251},
archivePrefix = {arXiv},
       eprint = {1601.01026},
 primaryClass = {astro-ph.GA},
       adsurl = {https://ui.adsabs.harvard.edu/abs/2016MNRAS.460.3610O}
}

@ARTICLE{slowstars,
       author = {{Shen}, Zili and {van Dokkum}, Pieter and {Danieli}, Shany},
        title = "{Confirmation of an Anomalously Low Dark Matter Content for the Galaxy NGC 1052-DF4 from Deep, High-resolution Continuum Spectroscopy}",
      journal = {\apj},
         year = 2023,
        month = nov,
       volume = {957},
       number = {1},
          eid = {6},
        pages = {6},
          doi = {10.3847/1538-4357/acfa70},
archivePrefix = {arXiv},
       eprint = {2309.08592},
 primaryClass = {astro-ph.GA},
       adsurl = {https://ui.adsabs.harvard.edu/abs/2023ApJ...957....6S}
}

@ARTICLE{grishin,
       author = {{Grishin}, Kirill A. and {Chilingarian}, Igor V. and {Afanasiev}, Anton V. and {Fabricant}, Daniel and {Katkov}, Ivan Yu. and {Moran}, Sean and {Yagi}, Masafumi},
        title = "{Transforming gas-rich low-mass disky galaxies into ultra-diffuse galaxies by ram pressure}",
      journal = {Nature Astronomy},
         year = 2021,
        month = dec,
       volume = {5},
        pages = {1308-1318},
          doi = {10.1038/s41550-021-01470-5},
archivePrefix = {arXiv},
       eprint = {2111.01140},
 primaryClass = {astro-ph.GA},
       adsurl = {https://ui.adsabs.harvard.edu/abs/2021NatAs...5.1308G}
}

@ARTICLE{benudgs,
       author = {{Benavides}, Jos{\'e} A. and {Sales}, Laura V. and {Abadi}, Mario. G. and {Pillepich}, Annalisa and {Nelson}, Dylan and {Marinacci}, Federico and {Cooper}, Michael and {Pakmor}, Ruediger and {Torrey}, Paul and {Vogelsberger}, Mark and {Hernquist}, Lars},
        title = "{Quiescent ultra-diffuse galaxies in the field originating from backsplash orbits}",
      journal = {Nature Astronomy},
         year = 2021,
        month = sep,
       volume = {5},
        pages = {1255-1260},
          doi = {10.1038/s41550-021-01458-1},
archivePrefix = {arXiv},
       eprint = {2109.01677},
 primaryClass = {astro-ph.GA},
       adsurl = {https://ui.adsabs.harvard.edu/abs/2021NatAs...5.1255B}
}

@ARTICLE{hartke,
       author = {{Hartke}, J. and {Iodice}, E. and {Gullieuszik}, M. and {Mirabile}, M. and {Buttitta}, C. and {Doll}, G. and {D'Ago}, G. and {de la Casa}, C.~C. and {Hess}, K.~M. and {Kotulla}, R. and {Poggianti}, B. and {Arnaboldi}, M. and {Cantiello}, M. and {Corsini}, E.~M. and {Falc{\'o}n-Barroso}, J. and {Forbes}, D.~A. and {Hilker}, M. and {Mieske}, S. and {Rejkuba}, M. and {Spavone}, M. and {Spiniello}, C.},
        title = "{Looking into the faintEst WIth MUSE (LEWIS): Exploring the nature of ultra-diffuse galaxies in the Hydra I cluster: III. Untangling UDG 32 from the stripped filaments of NGC 3314A with multi-wavelength data}",
      journal = {\aap},
         year = 2025,
        month = mar,
       volume = {695},
          eid = {A91},
        pages = {A91},
          doi = {10.1051/0004-6361/202452975},
archivePrefix = {arXiv},
       eprint = {2501.16192},
 primaryClass = {astro-ph.GA},
       adsurl = {https://ui.adsabs.harvard.edu/abs/2025A&A...695A..91H}
}

@ARTICLE{3DVirgo,
       author = {{Gavazzi}, G. and {Boselli}, A. and {Scodeggio}, M. and {Pierini}, D. and {Belsole}, E.},
        title = "{The 3D structure of the Virgo cluster from H-band Fundamental Plane and Tully-Fisher distance determinations}",
      journal = {\mnras},
         year = 1999,
        month = apr,
       volume = {304},
       number = {3},
        pages = {595-610},
          doi = {10.1046/j.1365-8711.1999.02350.x},
archivePrefix = {arXiv},
       eprint = {astro-ph/9812275},
 primaryClass = {astro-ph},
       adsurl = {https://ui.adsabs.harvard.edu/abs/1999MNRAS.304..595G}
}

@ARTICLE{barnes,
       author = {{Barnes}, D.~G. and {Staveley-Smith}, L. and {de Blok}, W.~J.~G. and {Oosterloo}, T. and {Stewart}, I.~M. and {Wright}, A.~E. and {Banks}, G.~D. and {Bhathal}, R. and {Boyce}, P.~J. and {Calabretta}, M.~R. and {Disney}, M.~J. and {Drinkwater}, M.~J. and {Ekers}, R.~D. and {Freeman}, K.~C. and {Gibson}, B.~K. and {Green}, A.~J. and {Haynes}, R.~F. and {te Lintel Hekkert}, P. and {Henning}, P.~A. and {Jerjen}, H. and {Juraszek}, S. and {Kesteven}, M.~J. and {Kilborn}, V.~A. and {Knezek}, P.~M. and {Koribalski}, B. and {Kraan-Korteweg}, R.~C. and {Malin}, D.~F. and {Marquarding}, M. and {Minchin}, R.~F. and {Mould}, J.~R. and {Price}, R.~M. and {Putman}, M.~E. and {Ryder}, S.~D. and {Sadler}, E.~M. and {Schr{\"o}der}, A. and {Stootman}, F. and {Webster}, R.~L. and {Wilson}, W.~E. and {Ye}, T.},
        title = "{The HI Parkes All Sky Survey: southern observations, calibration and robust imaging}",
      journal = {\mnras},
         year = 2001,
        month = apr,
       volume = {322},
       number = {3},
        pages = {486-498},
          doi = {10.1046/j.1365-8711.2001.04102.x},
       adsurl = {https://ui.adsabs.harvard.edu/abs/2001MNRAS.322..486B}
}

@ARTICLE{mei3Dvirgo,
       author = {{Mei}, Simona and {Blakeslee}, John P. and {C{\^o}t{\'e}}, Patrick and {Tonry}, John L. and {West}, Michael J. and {Ferrarese}, Laura and {Jord{\'a}n}, Andr{\'e}s and {Peng}, Eric W. and {Anthony}, Andr{\'e} and {Merritt}, David},
        title = "{The ACS Virgo Cluster Survey. XIII. SBF Distance Catalog and the Three-dimensional Structure of the Virgo Cluster}",
      journal = {\apj},
         year = 2007,
        month = jan,
       volume = {655},
       number = {1},
        pages = {144-162},
          doi = {10.1086/509598},
archivePrefix = {arXiv},
       eprint = {astro-ph/0702510},
 primaryClass = {astro-ph},
       adsurl = {https://ui.adsabs.harvard.edu/abs/2007ApJ...655..144M}
}

@ARTICLE{udgsnomolecules,
       author = {{Kado-Fong}, Erin and {Kim}, Chang-Goo and {Greene}, Jenny E. and {Lancaster}, Lachlan},
        title = "{Ultra-diffuse Galaxies as Extreme Star-forming Environments. II. Star Formation and Pressure Balance in H I-rich UDGs}",
      journal = {\apj},
         year = 2022,
        month = nov,
       volume = {939},
       number = {2},
          eid = {101},
        pages = {101},
          doi = {10.3847/1538-4357/ac9673},
archivePrefix = {arXiv},
       eprint = {2209.05500},
 primaryClass = {astro-ph.GA},
       adsurl = {https://ui.adsabs.harvard.edu/abs/2022ApJ...939..101K}
}

@ARTICLE{dwarfsco,
       author = {{Leroy}, A. and {Bolatto}, A.~D. and {Simon}, J.~D. and {Blitz}, L.},
        title = "{The Molecular Interstellar Medium of Dwarf Galaxies on Kiloparsec Scales: A New Survey for CO in Northern, IRAS-detected Dwarf Galaxies}",
      journal = {\apj},
         year = 2005,
        month = jun,
       volume = {625},
       number = {2},
        pages = {763-784},
          doi = {10.1086/429578},
archivePrefix = {arXiv},
       eprint = {astro-ph/0502302},
 primaryClass = {astro-ph},
       adsurl = {https://ui.adsabs.harvard.edu/abs/2005ApJ...625..763L}
}

@ARTICLE{virgodwco,
       author = {{Grossi}, M. and {Corbelli}, E. and {Bizzocchi}, L. and {Giovanardi}, C. and {Bomans}, D. and {Coelho}, B. and {De Looze}, I. and {Gon{\c{c}}alves}, T.~S. and {Hunt}, L.~K. and {Leonardo}, E. and {Madden}, S. and {Men{\'e}ndez-Delmestre}, K. and {Pappalardo}, C. and {Riguccini}, L.},
        title = "{Star-forming dwarf galaxies in the Virgo cluster: the link between molecular gas, atomic gas, and dust}",
      journal = {\aap},
         year = 2016,
        month = may,
       volume = {590},
          eid = {A27},
        pages = {A27},
          doi = {10.1051/0004-6361/201628123},
archivePrefix = {arXiv},
       eprint = {1602.09077},
 primaryClass = {astro-ph.GA},
       adsurl = {https://ui.adsabs.harvard.edu/abs/2016A&A...590A..27G}
}

@ARTICLE{almaram,
       author = {{Cramer}, W.~J. and {Kenney}, J.~D.~P. and {Cortes}, J.~R. and {Cortes P.~C.} and {Vlahakis}, C. and {J{\'a}chym}, P. and {Pompei}, E. and {Rubio}, M.},
        title = "{ALMA Evidence for Ram Pressure Compression and Stripping of Molecular Gas in the Virgo Cluster Galaxy NGC 4402}",
      journal = {\apj},
         year = 2020,
        month = oct,
       volume = {901},
       number = {2},
          eid = {95},
        pages = {95},
          doi = {10.3847/1538-4357/abaf54},
archivePrefix = {arXiv},
       eprint = {1910.14082},
 primaryClass = {astro-ph.GA},
       adsurl = {https://ui.adsabs.harvard.edu/abs/2020ApJ...901...95C}
}

@ARTICLE{rpsf,
       author = {{J{\'a}chym}, Pavel and {Kenney}, Jeffrey D.~P. and {Sun}, Ming and {Combes}, Fran{\c{c}}oise and {Cortese}, Luca and {Scott}, Tom C. and {Sivanandam}, Suresh and {Brinks}, Elias and {Roediger}, Elke and {Palou{\v{s}}}, Jan and {Fumagalli}, Michele},
        title = "{ALMA Unveils Widespread Molecular Gas Clumps in the Ram Pressure Stripped Tail of the Norma Jellyfish Galaxy}",
      journal = {\apj},
         year = 2019,
        month = oct,
       volume = {883},
       number = {2},
          eid = {145},
        pages = {145},
          doi = {10.3847/1538-4357/ab3e6c},
archivePrefix = {arXiv},
       eprint = {1905.13249},
 primaryClass = {astro-ph.GA},
       adsurl = {https://ui.adsabs.harvard.edu/abs/2019ApJ...883..145J}
}

@ARTICLE{yagitails,
       author = {{Yagi}, Masafumi and {Yoshida}, Michitoshi and {Komiyama}, Yutaka and {Kashikawa}, Nobunari and {Furusawa}, Hisanori and {Okamura}, Sadanori and {Graham}, Alister W. and {Miller}, Neal A. and {Carter}, David and {Mobasher}, Bahram and {Jogee}, Shardha},
        title = "{A Dozen New Galaxies Caught in the Act: Gas Stripping and Extended Emission Line Regions in the Coma Cluster}",
      journal = {\aj},
         year = 2010,
        month = dec,
       volume = {140},
       number = {6},
        pages = {1814-1829},
          doi = {10.1088/0004-6256/140/6/1814},
archivePrefix = {arXiv},
       eprint = {1005.3874},
 primaryClass = {astro-ph.CO},
       adsurl = {https://ui.adsabs.harvard.edu/abs/2010AJ....140.1814Y}
}

@ARTICLE{yagibig,
       author = {{Yagi}, Masafumi and {Yoshida}, Michitoshi and {Gavazzi}, Giuseppe and {Komiyama}, Yutaka and {Kashikawa}, Nobunari and {Okamura}, Sadanori},
        title = "{Extended Ionized Gas Clouds in the Abell 1367 Cluster}",
      journal = {\apj},
         year = 2017,
        month = apr,
       volume = {839},
       number = {1},
          eid = {65},
        pages = {65},
          doi = {10.3847/1538-4357/aa68e3},
archivePrefix = {arXiv},
       eprint = {1703.10301},
 primaryClass = {astro-ph.GA},
       adsurl = {https://ui.adsabs.harvard.edu/abs/2017ApJ...839...65Y}
}

@ARTICLE{morestellarmasses,
       author = {{Roediger}, Joel C. and {Courteau}, St{\'e}phane},
        title = "{On the uncertainties of stellar mass estimates via colour measurements}",
      journal = {\mnras},
         year = 2015,
        month = sep,
       volume = {452},
       number = {3},
        pages = {3209-3225},
          doi = {10.1093/mnras/stv1499},
archivePrefix = {arXiv},
       eprint = {1507.03016},
 primaryClass = {astro-ph.GA},
       adsurl = {https://ui.adsabs.harvard.edu/abs/2015MNRAS.452.3209R}
}

@ARTICLE{defdef,
       author = {{Giovanelli}, R. and {Haynes}, M.~P.},
        title = "{The HI extent and deficiency of spiral galaxies in the Virgo cluster.}",
      journal = {\aj},
         year = 1983,
        month = jul,
       volume = {88},
        pages = {881-908},
          doi = {10.1086/113376},
       adsurl = {https://ui.adsabs.harvard.edu/abs/1983AJ.....88..881G}
}

@ARTICLE{opticalTFR,
       author = {{Kourkchi}, Ehsan and {Tully}, R. Brent and {Anand}, Gagandeep S. and {Courtois}, H{\'e}l{\`e}ne M. and {Dupuy}, Alexandra and {Neill}, James D. and {Rizzi}, Luca and {Seibert}, Mark},
        title = "{Cosmicflows-4: The Calibration of Optical and Infrared Tully-Fisher Relations}",
      journal = {\apj},
         year = 2020,
        month = jun,
       volume = {896},
       number = {1},
          eid = {3},
        pages = {3},
          doi = {10.3847/1538-4357/ab901c},
archivePrefix = {arXiv},
       eprint = {2004.14499},
 primaryClass = {astro-ph.GA},
       adsurl = {https://ui.adsabs.harvard.edu/abs/2020ApJ...896....3K}
}

@ARTICLE{hidatabase,
       author = {{Courtois}, H{\'e}l{\`e}ne M. and {Tully}, R. Brent and {Fisher}, J. Richard and {Bonhomme}, Nicolas and {Zavodny}, Maximilian and {Barnes}, Austin},
        title = "{The Extragalactic Distance Database: All Digital H I Profile Catalog}",
      journal = {\aj},
         year = 2009,
        month = dec,
       volume = {138},
       number = {6},
        pages = {1938-1956},
          doi = {10.1088/0004-6256/138/6/1938},
archivePrefix = {arXiv},
       eprint = {0902.3670},
 primaryClass = {astro-ph.CO},
       adsurl = {https://ui.adsabs.harvard.edu/abs/2009AJ....138.1938C}
}

@ARTICLE{anotherhud,
       author = {{Papastergis}, E. and {Adams}, E.~A.~K. and {Romanowsky}, A.~J.},
        title = "{The HI content of isolated ultra-diffuse galaxies: A sign of multiple formation mechanisms?}",
      journal = {\aap},
         year = 2017,
        month = may,
       volume = {601},
          eid = {L10},
        pages = {L10},
          doi = {10.1051/0004-6361/201730795},
archivePrefix = {arXiv},
       eprint = {1703.05610},
 primaryClass = {astro-ph.GA},
       adsurl = {https://ui.adsabs.harvard.edu/abs/2017A&A...601L..10P}
}

@ARTICLE{smudgeshi,
       author = {{Karunakaran}, Ananthan and {Motiwala}, Khadeejah and {Spekkens}, Kristine and {Zaritsky}, Dennis and {Donnerstein}, Richard L. and {Dey}, Arjun},
        title = "{Systematically Measuring Ultradiffuse Galaxies. VII. The H I Survey Overview}",
      journal = {\apj},
         year = 2024,
        month = nov,
       volume = {975},
       number = {1},
          eid = {91},
        pages = {91},
          doi = {10.3847/1538-4357/ad77cf},
archivePrefix = {arXiv},
       eprint = {2408.07119},
 primaryClass = {astro-ph.GA},
       adsurl = {https://ui.adsabs.harvard.edu/abs/2024ApJ...975...91K}
}

@ARTICLE{jones23,
       author = {{Jones}, Michael G. and {Karunakaran}, Ananthan and {Bennet}, Paul and {Sand}, David J. and {Spekkens}, Kristine and {Mutlu-Pakdil}, Bur{\c{c}}in and {Crnojevi{\'c}}, Denija and {Janowiecki}, Steven and {Leisman}, Lukas and {Fielder}, Catherine E.},
        title = "{Gas-rich, Field Ultra-diffuse Galaxies Host Few Globular Clusters}",
      journal = {\apjl},
         year = 2023,
        month = jan,
       volume = {942},
       number = {1},
          eid = {L5},
        pages = {L5},
          doi = {10.3847/2041-8213/acaaab},
archivePrefix = {arXiv},
       eprint = {2211.00651},
 primaryClass = {astro-ph.GA},
       adsurl = {https://ui.adsabs.harvard.edu/abs/2023ApJ...942L...5J}
}

@ARTICLE{forbes,
       author = {{Forbes}, Duncan A. and {Alabi}, Adebusola and {Romanowsky}, Aaron J. and {Brodie}, Jean P. and {Arimoto}, Nobuo},
        title = "{Globular clusters in Coma cluster ultra-diffuse galaxies (UDGs): evidence for two types of UDG?}",
      journal = {\mnras},
         year = 2020,
        month = mar,
       volume = {492},
       number = {4},
        pages = {4874-4883},
          doi = {10.1093/mnras/staa180},
archivePrefix = {arXiv},
       eprint = {2001.10031},
 primaryClass = {astro-ph.GA},
       adsurl = {https://ui.adsabs.harvard.edu/abs/2020MNRAS.492.4874F}
}

@ARTICLE{virgoudgs,
       author = {{Lim}, Sungsoon and {C{\^o}t{\'e}}, Patrick and {Peng}, Eric W. and {Ferrarese}, Laura and {Roediger}, Joel C. and {Durrell}, Patrick R. and {Mihos}, J. Christopher and {Wang}, Kaixiang and {Gwyn}, S.~D.~J. and {Cuillandre}, Jean-Charles and {Liu}, Chengze and {S{\'a}nchez-Janssen}, Rub{\'e}n and {Toloba}, Elisa and {Sales}, Laura V. and {Guhathakurta}, Puragra and {Lan{\c{c}}on}, Ariane and {Puzia}, Thomas H.},
        title = "{The Next Generation Virgo Cluster Survey (NGVS). XXX. Ultra-diffuse Galaxies and Their Globular Cluster Systems}",
      journal = {\apj},
         year = 2020,
        month = aug,
       volume = {899},
       number = {1},
          eid = {69},
        pages = {69},
          doi = {10.3847/1538-4357/aba433},
archivePrefix = {arXiv},
       eprint = {2007.10565},
 primaryClass = {astro-ph.GA},
       adsurl = {https://ui.adsabs.harvard.edu/abs/2020ApJ...899...69L}
}

@ARTICLE{tailwidth1,
       author = {{Fossati}, Matteo and {Fumagalli}, Michele and {Boselli}, Alessandro and {Gavazzi}, Giuseppe and {Sun}, Ming and {Wilman}, David J.},
        title = "{MUSE sneaks a peek at extreme ram-pressure stripping events - II. The physical properties of the gas tail of ESO137-001}",
      journal = {\mnras},
         year = 2016,
        month = jan,
       volume = {455},
       number = {2},
        pages = {2028-2041},
          doi = {10.1093/mnras/stv2400},
archivePrefix = {arXiv},
       eprint = {1510.04283},
 primaryClass = {astro-ph.GA},
       adsurl = {https://ui.adsabs.harvard.edu/abs/2016MNRAS.455.2028F}
}

@ARTICLE{tailwidth2,
       author = {{Ignesti}, Alessandro and {Brunetti}, Gianfranco and {Gullieuszik}, Marco and {Akerman}, Nina and {Marasco}, Antonino and {Poggianti}, Bianca M. and {Li}, Yuan and {Vulcani}, Benedetta and {Gitti}, Myriam and {Moretti}, Alessia and {Giunchi}, Eric and {Tomi{\v{c}}i{\'c}}, Neven and {Bacchini}, Cecilia and {Paladino}, Rosita and {Radovich}, Mario and {Wolter}, Anna},
        title = "{Investigating the Intracluster Medium Viscosity Using the Tails of GASP Jellyfish Galaxies}",
      journal = {\apj},
         year = 2024,
        month = dec,
       volume = {977},
       number = {2},
          eid = {219},
        pages = {219},
          doi = {10.3847/1538-4357/ad919b},
archivePrefix = {arXiv},
       eprint = {2411.07034},
 primaryClass = {astro-ph.CO},
       adsurl = {https://ui.adsabs.harvard.edu/abs/2024ApJ...977..219I}
}

@ARTICLE{tailwidth3,
       author = {{Kenney}, Jeffrey D.~P. and {van Gorkom}, J.~H. and {Vollmer}, B.},
        title = "{VLA H I Observations of Gas Stripping in the Virgo Cluster Spiral NGC 4522}",
      journal = {\aj},
         year = 2004,
        month = jun,
       volume = {127},
       number = {6},
        pages = {3361-3374},
          doi = {10.1086/420805},
archivePrefix = {arXiv},
       eprint = {astro-ph/0403103},
 primaryClass = {astro-ph},
       adsurl = {https://ui.adsabs.harvard.edu/abs/2004AJ....127.3361K}
}

@ARTICLE{tailwidth4,
       author = {{Luo}, Rongxin and {Sun}, Ming and {J{\'a}chym}, Pavel and {Waldron}, Will and {Fossati}, Matteo and {Fumagalli}, Michele and {Boselli}, Alessandro and {Combes}, Francoise and {Kenney}, Jeffrey D.~P. and {Li}, Yuan and {Gronke}, Max},
        title = "{Tracing the kinematics of the whole ram-pressure-stripped tails in ESO 137-001}",
      journal = {\mnras},
         year = 2023,
        month = jun,
       volume = {521},
       number = {4},
        pages = {6266-6283},
          doi = {10.1093/mnras/stad1003},
archivePrefix = {arXiv},
       eprint = {2212.03891},
 primaryClass = {astro-ph.GA},
       adsurl = {https://ui.adsabs.harvard.edu/abs/2023MNRAS.521.6266L}
}

@ARTICLE{DF44_1,
       author = {{van Dokkum}, Pieter and {Abraham}, Roberto and {Brodie}, Jean and {Conroy}, Charlie and {Danieli}, Shany and {Merritt}, Allison and {Mowla}, Lamiya and {Romanowsky}, Aaron and {Zhang}, Jielai},
        title = "{A High Stellar Velocity Dispersion and {\ensuremath{\sim}}100 Globular Clusters for the Ultra-diffuse Galaxy Dragonfly 44}",
      journal = {\apjl},
         year = 2016,
        month = sep,
       volume = {828},
       number = {1},
          eid = {L6},
        pages = {L6},
          doi = {10.3847/2041-8205/828/1/L6},
archivePrefix = {arXiv},
       eprint = {1606.06291},
 primaryClass = {astro-ph.GA},
       adsurl = {https://ui.adsabs.harvard.edu/abs/2016ApJ...828L...6V}
}

@ARTICLE{DF44_2,
       author = {{Saifollahi}, Teymoor and {Trujillo}, Ignacio and {Beasley}, Michael A. and {Peletier}, Reynier F. and {Knapen}, Johan H.},
        title = "{The number of globular clusters around the iconic UDG DF44 is as expected for dwarf galaxies}",
      journal = {\mnras},
         year = 2021,
        month = apr,
       volume = {502},
       number = {4},
        pages = {5921-5934},
          doi = {10.1093/mnras/staa3016},
archivePrefix = {arXiv},
       eprint = {2006.14630},
 primaryClass = {astro-ph.GA},
       adsurl = {https://ui.adsabs.harvard.edu/abs/2021MNRAS.502.5921S}
}

@ARTICLE{vcc1287,
       author = {{Beasley}, Michael A. and {Romanowsky}, Aaron J. and {Pota}, Vincenzo and {Navarro}, Ignacio Martin and {Martinez Delgado}, David and {Neyer}, Fabian and {Deich}, Aaron L.},
        title = "{An Overmassive Dark Halo around an Ultra-diffuse Galaxy in the Virgo Cluster}",
      journal = {\apjl},
         year = 2016,
        month = mar,
       volume = {819},
       number = {2},
          eid = {L20},
        pages = {L20},
          doi = {10.3847/2041-8205/819/2/L20},
archivePrefix = {arXiv},
       eprint = {1602.04002},
 primaryClass = {astro-ph.GA},
       adsurl = {https://ui.adsabs.harvard.edu/abs/2016ApJ...819L..20B}
}

\end{document}